\documentclass[aps,pre,twocolumn,galleyproof,superscriptaddress,groupedaddress,showpacs]{revtex4}
\pdfoutput=1
\usepackage{graphicx,dsfont,amssymb}
\begin{document}
\title{Public transport networks: empirical  analysis and modeling}
\author{C.\ von Ferber}
\email[]{C.vonFerber@coventry.ac.uk}
 \affiliation{Applied Mathematics Research Centre, Coventry University,
Coventry CV1 5FB, UK}
 \affiliation{Theoretische Polymerphysik,
Universit\"at Freiburg, D-79104 Freiburg, Germany}
\author{T. Holovatch}\email[]{holtaras@lpm.u-nancy.fr}
\affiliation{Laboratoire de Physique des Mat\'eriaux,
               Universit\'e Henri Poincar\'e, Nancy 1,
               54506 Vand\oe uvre les Nancy Cedex, France}
 \affiliation{Applied Mathematics Research Centre, Coventry University,
Coventry CV1 5FB, UK}
\author{Yu. Holovatch}\email[]{hol@icmp.lviv.ua}
\affiliation{Institute for Condensed Matter Physics, National
Academy of Sciences of Ukraine, UA--79011 Lviv, Ukraine}
\affiliation{Institut f\"ur Theoretische Physik, Johannes Kepler
Universit\"at Linz, A-4040, Linz, Austria}
\author{V. Palchykov}\email[]{palchykov@icmp.lviv.ua}
\affiliation{Institute for Condensed Matter Physics, National
Academy of Sciences of Ukraine, UA--79011 Lviv, Ukraine}
\date{\today}
\begin{abstract}
We  use complex network concepts to analyze  statistical
properties of urban public transport networks (PTN). To this
end, we present a comprehensive survey of the  statistical
properties of PTNs based on the data of fourteen cities of so far
unexplored network size. Especially helpful in our analysis are
different network representations. Within a comprehensive
approach we calculate PTN characteristics in all of these representations
and perform a comparative analysis. The standard network characteristics
obtained in this way often correspond to features that are of
practical importance to a passenger using public traffic in a given
city. Specific features are addressed that are unique
to PTNs and networks with similar transport functions (such as
networks of neurons, cables, pipes, vessels embedded in 2D or 3D
space). Based on the empirical survey, we propose a model that
albeit being simple enough is capable of reproducing many of the
identified PTN properties.
A central ingredient of this model is a growth dynamics in
terms of routes represented by self-avoiding walks.
\end{abstract}
\pacs{02.50.-r, 07.05.Rm, 89.75.Hc}
\maketitle
\section{Introduction}\label{I}

The general interest in networks of man-made and natural systems has
lead to a careful analysis of various network instances using
empirical, simulational, and theoretical tools. The emergence of this field
is sometimes referred to as the birth of network science
\cite{Barabasi02,Dorogovtsev02,Newman03a,Dorogovtsev03,Holovatch06}.
In this paper, we use complex network concepts to analyze the
statistical properties of public transport networks (PTN) of large cities.
These constitute an example of transportation networks
\cite{Newman03a} and share general features of these systems:
evolutionary dynamics, optimization, embedding in two dimensional (2D)
 space. Other
examples of transportation networks are given by the airport
\cite{Amaral00,Guimera04,Guimera05,Barrat04,Chi03,He04,Li04,Li06},
railway \cite{Sen03}, or power grid networks
\cite{Amaral00,Crucitti04,Albert04}.

\begin{figure}[ht]
 \centerline{\includegraphics[width=8cm]{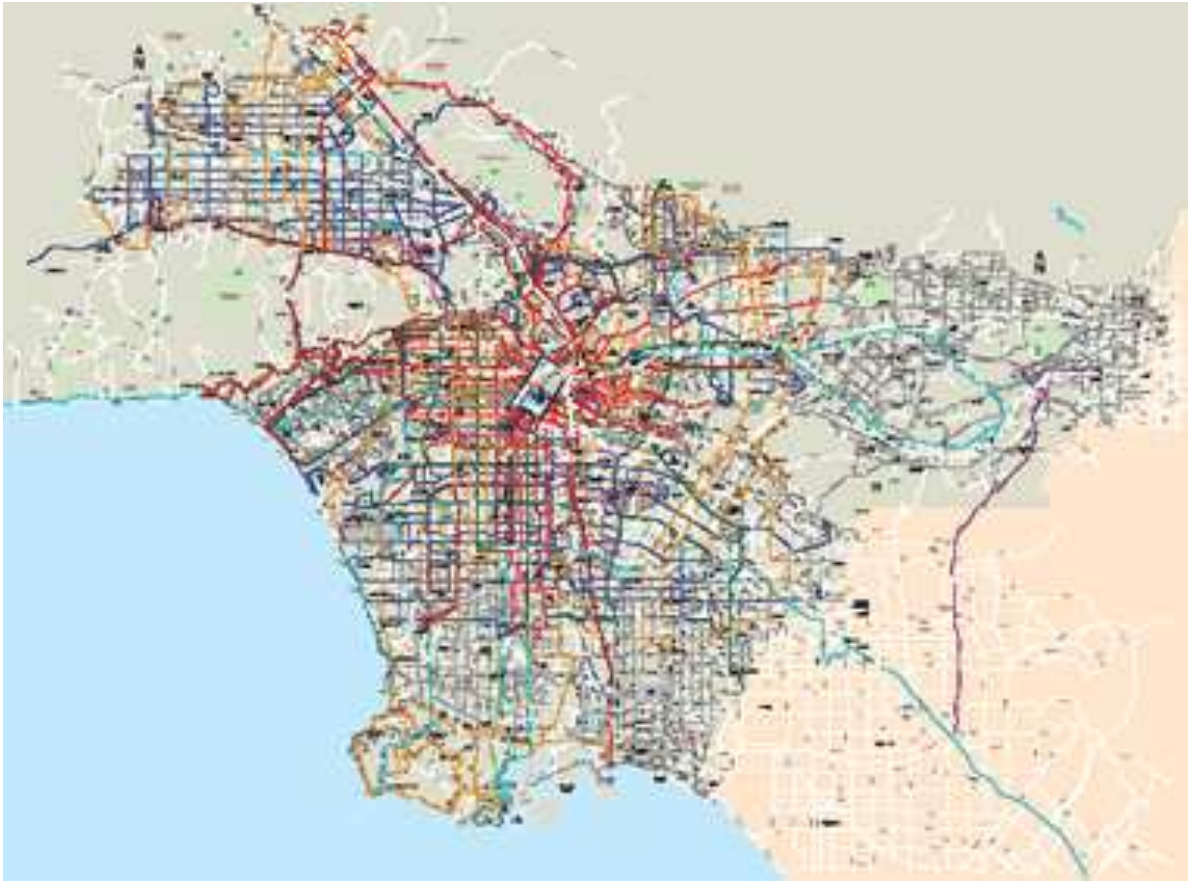}}
   \caption{One of the networks we analyze in this study. The Los Angeles PTN
   consists of $R=1881$ routes and $N=44629$
   stations, some of them are shown in this map (color online).
   \label{fig1}}
\end{figure}

While the evolution of a  PTN of a given city is closely related to
the city growth itself and therefore is influenced by numerous
factors of geographical, historical, and social origin, there is
ample evidence that PTNs of different cities share common
statistical properties that arise due to their functional purposes
\cite{Marchiori00,Latora01,Latora02,Seaton04,Ferber05,Sienkiewicz05,Angeloudis06,
Zhang06,Ferber07a,Chang07,Xu07,Ferber07b,Ferber07c}. Some of these
properties have been analyzed in former studies, however, the objective of
the present study is to present a comprehensive survey of
 characteristics of PTNs and to provide a comparative analysis. Based
on this empirical survey we are in the position to propose
a growth model that captures many of the main (statistical) features of PTNs.

A further distinct feature of our study is that the PTNs we will
consider are networks of {\em all} means of public transport of
a city (buses, trams, subway, etc.) regardless of the specific means
of transport. A number of studies have analyzed specific
sub-networks of PTNs
\cite{Marchiori00,Latora01,Latora02,Seaton04,Angeloudis06,Zhang06,Xu07}.
Examples are the Boston
\cite{Marchiori00,Latora01,Latora02,Seaton04} and  Vienna
\cite{Seaton04} subway networks and the bus networks of several
cities in China \cite{Zhang06,Xu07}. However, each particular
traffic system (e.g. the network of buses or trams, or the subway
network) is not a closed system: it is a subgraph of a wider
transportation system of a city, or as we call it here, of a PTN.
Therefore to understand and describe the properties of public transport in
a city as a whole, one should analyze the complete network, without
restriction to specific parts. Indeed, for the case of Boston it has
been shown that changing from the subway system to the network ``subway
+ bus" the network properties change drastically
\cite{Latora01,Latora02}.

Urban public transport networks of general type have so far been
analyzed mainly in two previous studies \cite{Ferber05,Sienkiewicz05}. In
the first one, Ref. \cite{Ferber05}, the PTNs of Berlin,
D\"usseldorf, and Paris were examined, whereas the subject of Ref.
\cite{Sienkiewicz05} were public transport systems of 22 Polish
cities. Ref. \cite{Ferber05} concentrated on the scale-free
properties. For the cities considered, the node degree distribution
was shown to follow a power law. Moreover power laws were found for
a number of other specific features describing the traffic load on the
PTN. However, the statistics in this study was too small for definite
conclusions. In Ref. \cite{Sienkiewicz05} it was found that the node
degree distribution may follow a power law or be described by an
exponential function, depending on the assumed network representation.
 Besides, a number of other network characteristics
(clustering, betweenness, assortativity) were extensively analyzed.

In the present paper, we analyze PTNs of a number of major cities of
the world (see table \ref{tab1}) \cite{database1,database2}. Our
choice for this data base was motivated by the requirement to collect
network samples from cities of different  geographical,
cultural, and economical background. Our current analysis
extends former studies \cite{Ferber05,Sienkiewicz05} by
considering cities with larger public transport systems (the typical
number of stops in the systems considered in Ref.
\cite{Sienkiewicz05} was several hundreds) as well as by systematically
analyzing different representations. The idea of different
network representations naturally arises in the network science
\cite{Barabasi02,Dorogovtsev02,Newman03a,Dorogovtsev03,Holovatch06}.
For the PTN the primary network topology is given by the set of
routes each servicing an ordered series of given stations (see Fig.
\ref{fig1} as an example). For the transportation networks studied
so far mainly two different neighborhood relations were used. In the first
one, two stations are defined as neighbors only if one station is
the successor of the other in the series serviced by this route
\cite{Latora01,Latora02}. In the second one, two stations are
neighbors whenever they are serviced by a common route \cite{Sen03}.
We will exploit both representations in our study. Moreover, we introduce
further natural representations (described in detail in Section
\ref{II}) which make the description of the PTNs of table \ref{tab1}
comprehensive. In particular, this includes a bipartite graph representation of
a transportation network  that reflects its intrinsic features
\cite{Seaton04,Zhang06,Chang07}.

\begin{table}[htbp]
\centering
\tabcolsep1.5mm
\begin{tabular}{lrrrrrl}
\hline \hline City& $A$    &  $P$ &    $N$ & $R$    & $S$  & Type   \\ \hline
 Berlin        &      892  &  3.7  &  2992 &   211  & 29.4 & BSTU   \\
 Dallas        &      887  &  1.2  &  5366 &   117  & 59.9 & B      \\
 D\"usseldorf  &      217  &  0.6  &  1494 &   124  & 28.5 & BST    \\
 Hamburg       &      755  &  1.8  &  8084 &   708  & 25.5 & BFSTU  \\
 Hong Kong     &     1052  &  7.0  &  2024 &   321  & 39.6 & B      \\
 Istanbul      &     1538  &  11.1 &  4043 &   414  & 31.7 & BST    \\
 London        &     1577  &  8.3  &  10937&   922  & 34.2 & BST    \\
 Los Angeles   &     1214  &  3.8  &  44629&  1881  & 52.9 & B      \\
 Moscow        &     1081  & 10.5  &  3569 &   679  & 22.2 & BEST   \\
 Paris         &      2732 &  10.0 &  3728 &   251  & 38.2 & BS     \\
 Rome          &     5352  &  4.0  &  3961 &   681  & 26.8 & BT     \\
 Sa\~o Paolo   &     1523  & 10.9  &  7215 &   997  & 58.3 & B      \\
 Sydney        &     1687  &  3.6  &  1978 &   596  & 16.3 & B      \\
 Taipei        &     2457  &  6.8  &  5311 &   389  & 70.5 & B      \\
 \hline \hline
\end{tabular}
\caption{Cities analyzed in this study.
 $A$: urban area
(km$^2$);
 $P$: population (million inhabitants); $N$: number of PTN stations;
 $R$: number of PTN routes; $S$: mean route length. Type of transport
 which is taken into account in the PTN database:
 Bus, Electric trolleybus, Ferry, Subway, Tram, Urban train.}
\label{tab1}
\end{table}

There is another reason to seek scale-free properties of PTNs
considering a larger data base of more cities with larger public
transport communications involved. A currently well accepted
mechanism to explain the abundant occurrence of power laws is that
of preferential attachment  or ``rich gets richer"
\cite{Simon55,Price76,prefattach}. As far as PTNs obviously are
evolving networks, their evolution may be expected to
follow a similar underlying mechanisms. However, scale-free networks have also been
shown to arise when minimizing both the effort for communication and
the cost for maintaining connections \cite{optimization,Gastner04}.
Moreover, this kind of an optimization was shown to lead to small
world properties \cite{Mathias01} and to explain the appearance of
power laws in a general context \cite{zipfoptimal}. Therefore,
scale-free behavior of PTNs may also be related to obvious objectives
to optimize their operation.

This paper is organized as follows. In the next section (\ref{II}) we
define different representations in which the PTN will be
analyzed, sections \ref{III}-\ref{V} explore the network
properties in these representations.
We separately analyze local characteristics, such
as node degrees and clustering coefficients (section \ref{III}), and
global characteristics, such as path length distributions and
centralities (section \ref{IV}). Special attention is paid to
characteristics that are unique to PTNs and networks with similar
construction principles. An example is given by the analysis of
sequences of routes which go in parallel along a given sequence of
stations, a feature we call 'harness' effect.
A description of correlations
between the properties of neighboring nodes in terms of generalized
assortativities is performed in section \ref{V}. Our findings for
the statistics of real-word PTNs are supported by simulations of an
evolutionary model of PTNs as displayed in section \ref{VI}.
Conclusions and an outlook are given in section \ref{VII}. Some of
our results have been preliminary announced in Ref.
\cite{Ferber07a}.

\section{PT network topology}\label{II}

\begin{figure}[!h]
\tabcolsep3mm
\begin{tabular}{cc}
 \includegraphics[width=3.5cm]{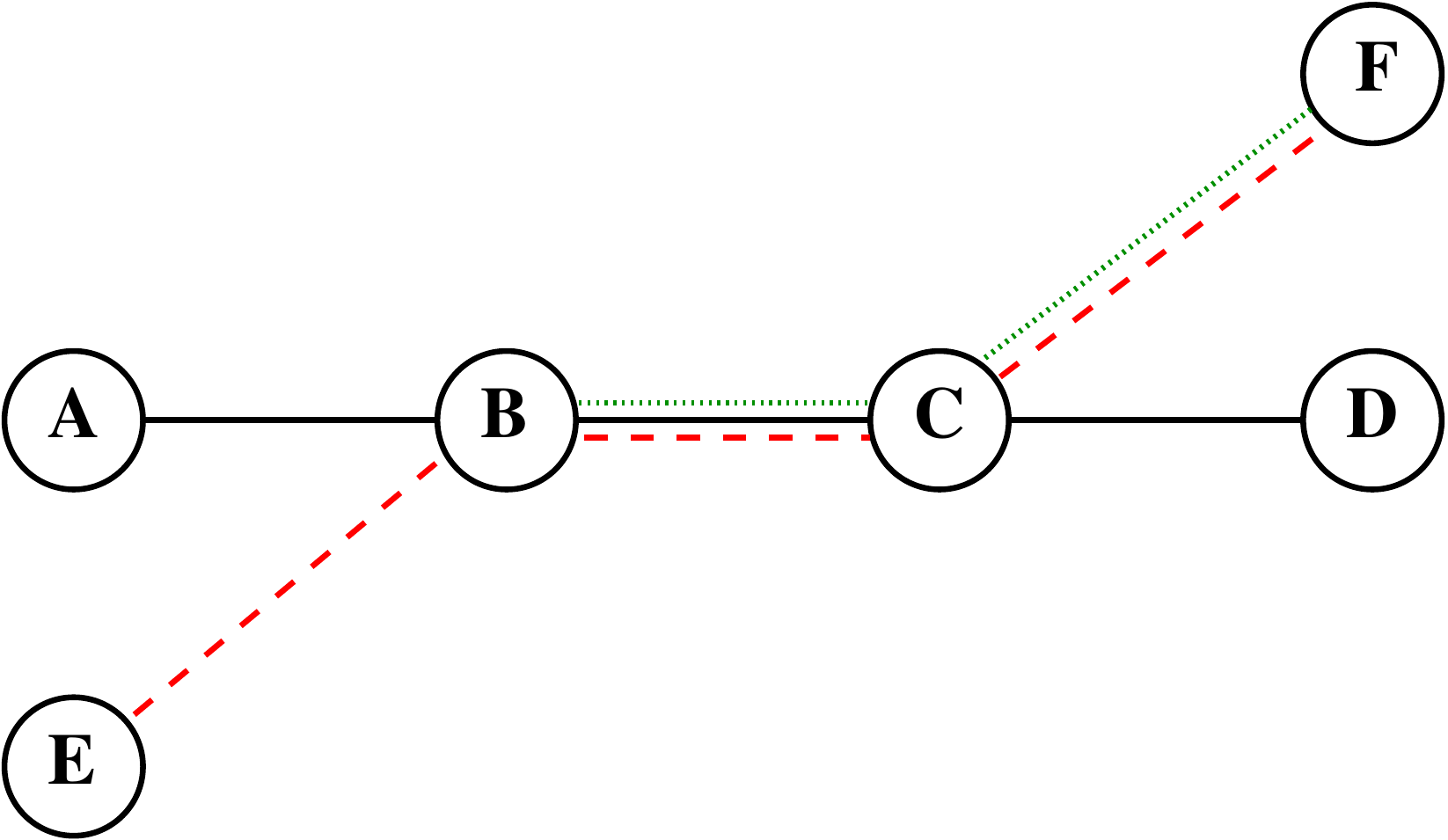} &
  \includegraphics[width=3.5cm]{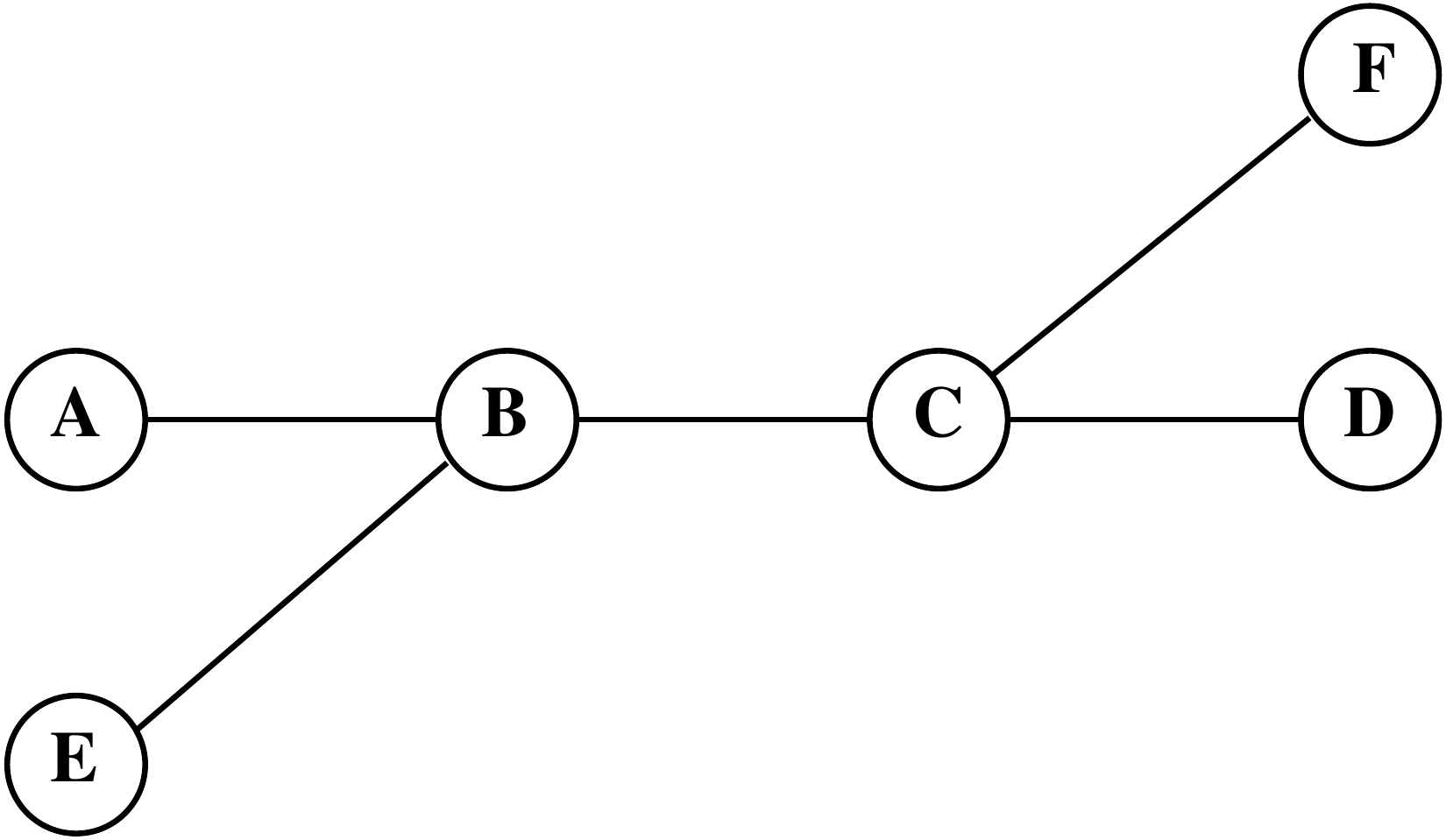} \\
  {\bf  a}  & {\bf  b} \\ \\
 \includegraphics[width=16mm]{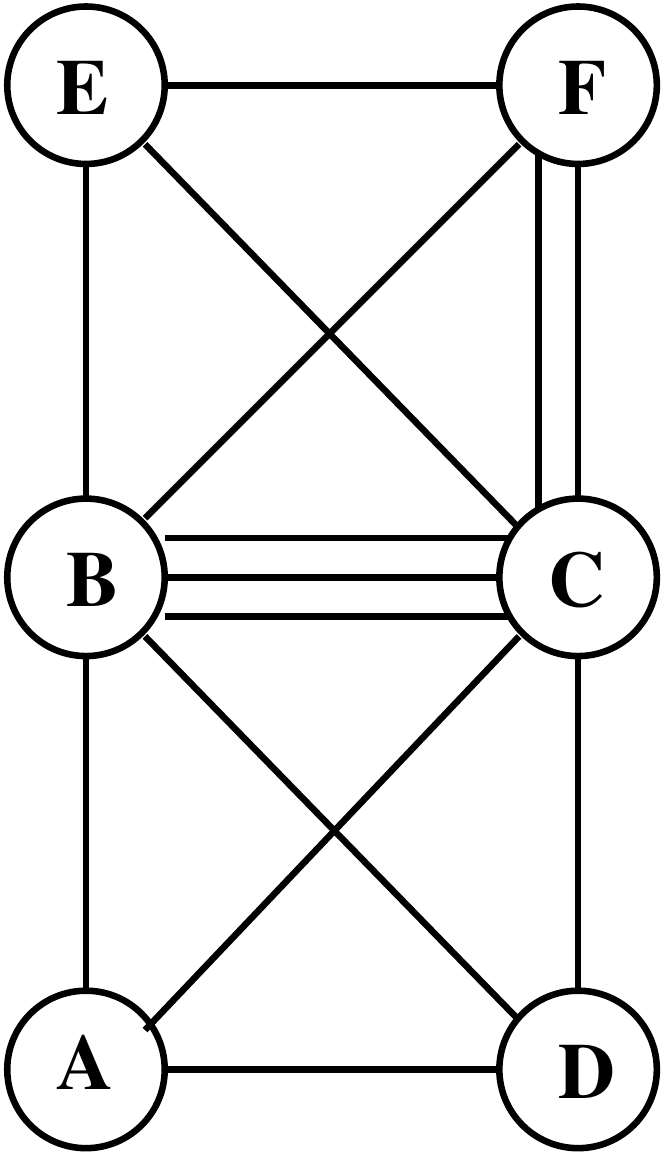}  &
  \includegraphics[width=16mm]{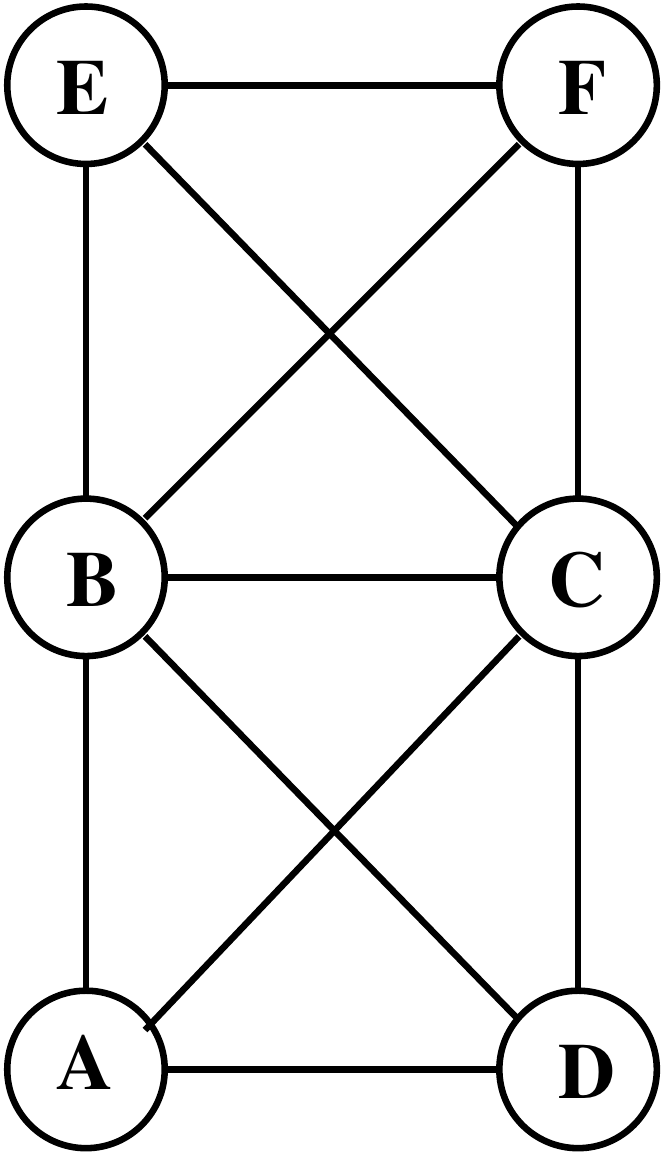}  \\
  {\bf  c}  & {\bf  d} \\ \\
   \includegraphics[width=2cm]{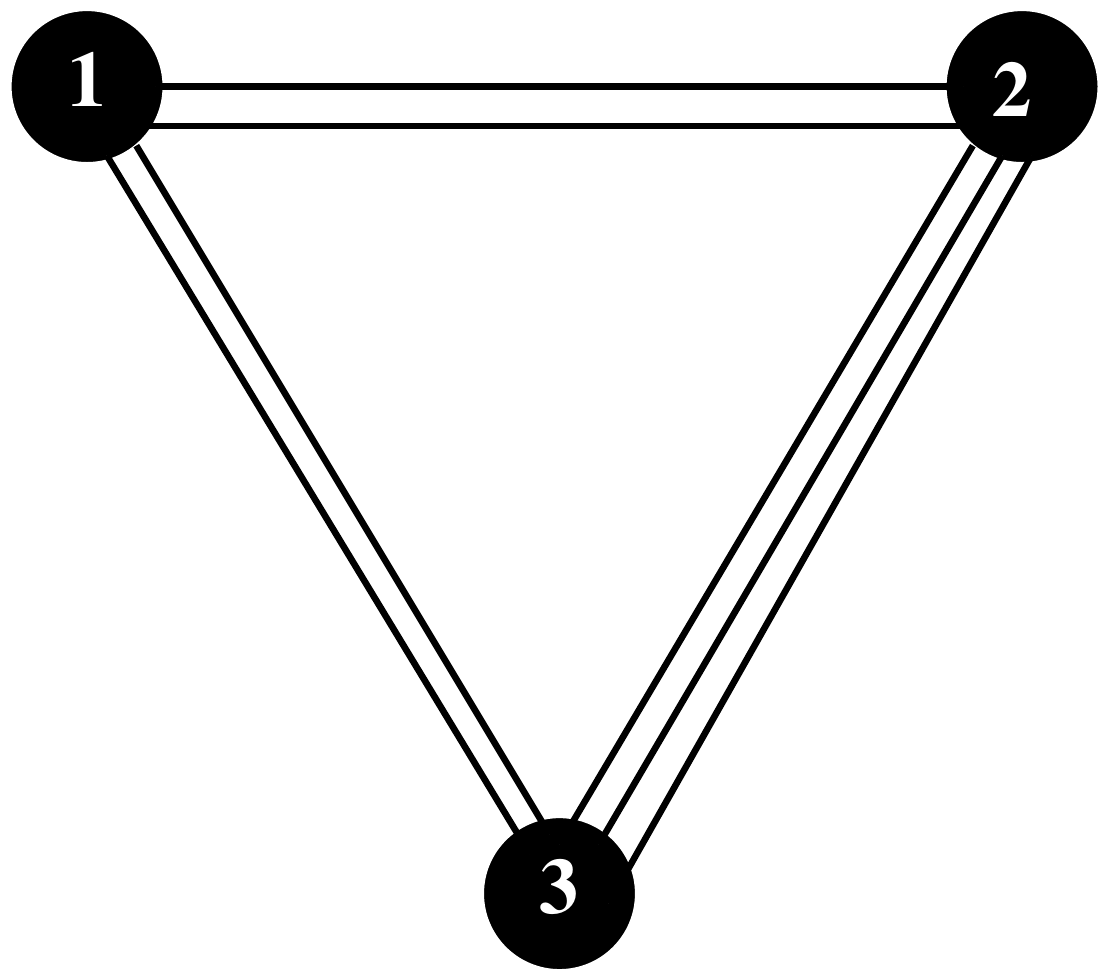} &
  \includegraphics[width=2cm]{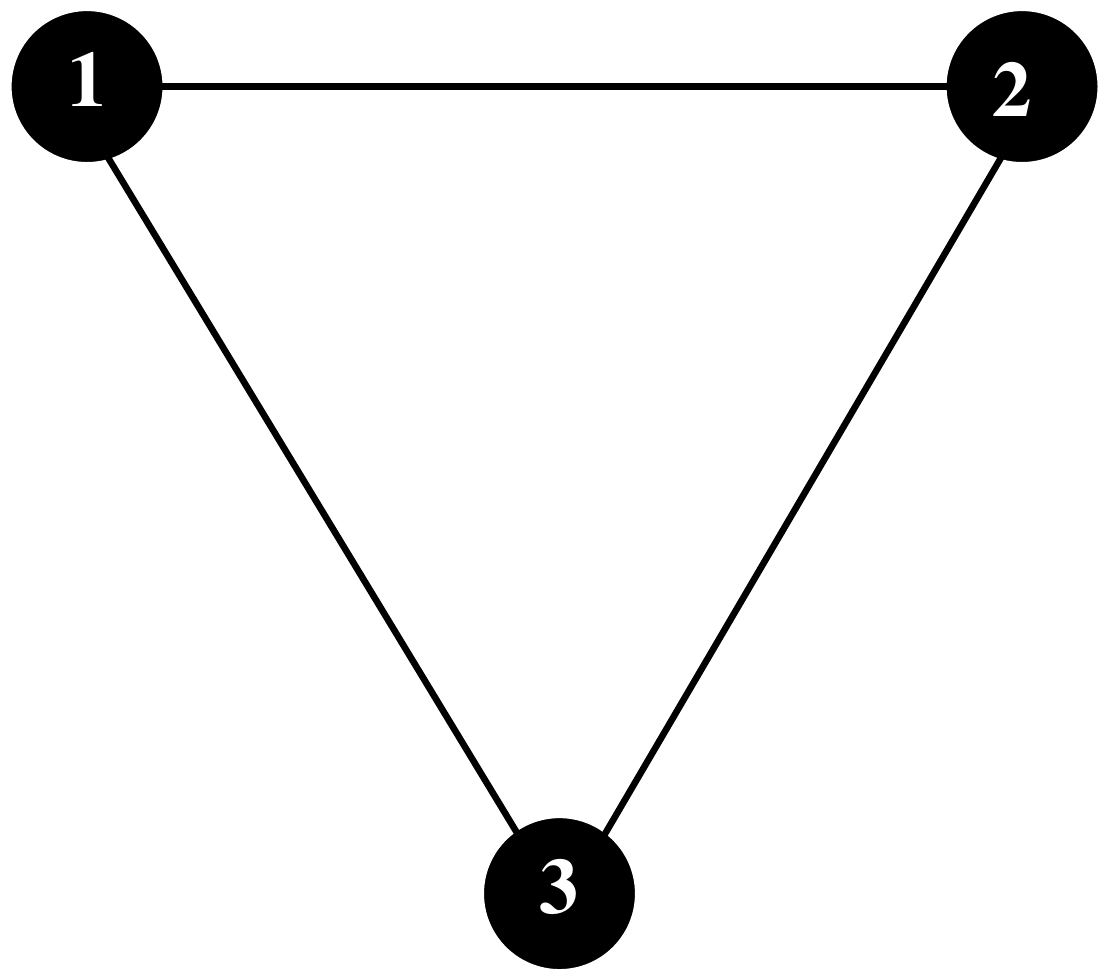} \\
  {\bf  e}  & {\bf  f} \\ \\
  \end{tabular}
   \caption{{\bf a}: a piece of public transport map. Stations A-F are serviced
   by the tram lines  No 1 (solid line), No 2 (dashed line),  No 3 (dotted line).
   Taking all the lines to be indistinguishable, we call such representation
   $\mathds{L}'$-space. {\bf b}: $\mathds{L}$-space. {\bf c}: $\mathds{P}'$-space.
   All stations that belong to the same route are connected.
   {\bf d}: $\mathds{P}$-space. {\bf e}: $\mathds{C}'$-space. Each route is represented
   by a node, each link corresponds to a common station shared by the route nodes it connects.
   {\bf f}: $\mathds{C}$-space.\label{fig2}}
\end{figure}

Although everyone has an intuitive idea about what a PTN is, it
appears that there are numerous ways to define its topology. Let us
describe some of them, defining different 'spaces' in which public
transport networks will be analyzed. A straightforward
representation of a PT map in the form of a graph
represents every station by a node while any two nodes
that are successively serviced by at least one route are linked by an edge as
shown in Fig. {\ref{fig2}\bf a}. Let us note, that the full
information about the network of $N$ stations and $R$ routes is
given by the set of ordered lists each corresponding to one route or
to one of the two directions of a given route. These simply list all stations
serviced by that route in the order of service between two terminal
stations or in the course of a round trip. Note that multiple
entries of a given station in such a list are possible and do
occur. Let us first introduce a simple graph to represent this
situation. In the following we will refer to this graph as a
$\mathds{L}$-space \cite{Sienkiewicz05}. This graph represents each
station by a node, a link between nodes indicates that there is at
least one route that services the corresponding station
consecutively. No multiple links are allowed (see Fig.
{\ref{fig2}\bf b}). The neighbors of a given node in
$\mathds{L}$-space represent all stations that are within reach of a
single station trip. For analyzing PTNs, the $\mathds{L}$-space
representation has been used in Refs.
\cite{Latora01,Ferber05,Sienkiewicz05, Angeloudis06,Xu07}. Extending
the notion of $\mathds{L}$-space one may either introduce multiple
links between nodes depending on the number of services between them
or associate a corresponding weight to a single link. We will refer
to such a representation as $\mathds{L}'$-space (c.f. Fig.
\ref{fig2}{\bf a}).

A particularly useful
concept for the description of connectivity in transport networks
which we refer to as $\mathds{P}$-space  \cite{Sienkiewicz05} was introduced
in ref. \cite{Sen03} and used in PTN analysis in Refs. \cite{Seaton04,Sienkiewicz05,Xu07}.
In this representation the network is a graph where stations are represented by nodes
that are linked if they are serviced by at least one common route. In
$\mathds{P}$-space representation
the neighborhood of a given node represents all stations
that can be reached without changing means of transport. The
$\mathds{P}$-space concept may be extended to include multiple or
weighted links. Such a representation we refer to as
$\mathds{P}'$-space (c.f. Figs. {\ref{fig2}\bf c} and {\ref{fig2}\bf
d}, correspondingly).

\begin{figure}[ht]

 \centerline{\includegraphics[width=8cm]{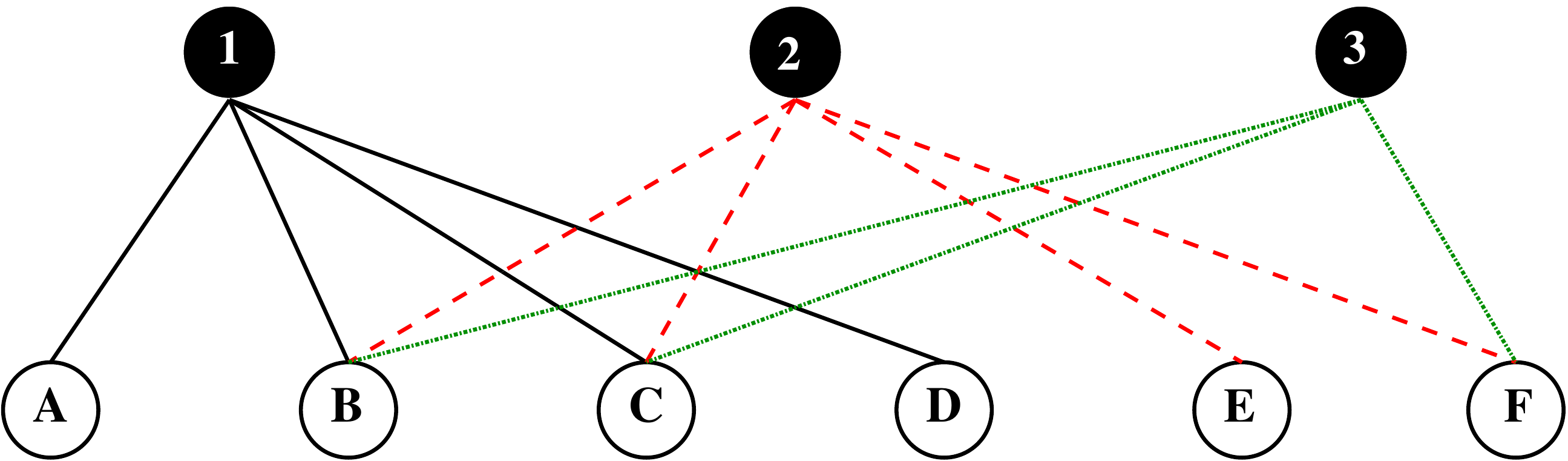}}
   \caption{A bipartite graph of  tram lines (filled circles) and
   stations (circles) which corresponds to the public transport map
   of Fig. {\ref{fig2}\bf a}. For the sake of illustration, lines
   corresponding to different tram routes are shown in a differing
   way. However, neither line type nor the order of the stations matter in this
   graph. Note that Figs. {\ref{fig2}\bf c} - {\ref{fig2}\bf f}
   are the one-mode projections of this  bipartite graph. \label{fig3}}
\end{figure}

A somewhat different concept is that of a bipartite space which is
useful in the analysis of cooperation networks \cite{Newman03a,Guillaume06}. In
this representation which we call $\mathds{B}$-space both routes and
stations are represented by nodes \cite{Ferber07a,Zhang06,Chang07}.
Each route node is linked
to all station nodes that it services. No direct links between nodes
of same type occur (see Fig. \ref{fig3}). Obviously, in
$\mathds{B}$-space the neighbors of a given route node are all
stations that it services while the neighbors of a given station
node are all routes that service it.

We note that the one mode projections of the bipartite graph of
$\mathds{B}$-space to the set of station nodes results in
$\mathds{P}$-space or in $\mathds{P}'$-space space if we retain
multiple links. The complementary projection to route nodes leads
to a graph which we call $\mathds{C}$-space ($\mathds{C}'$-space if
multiple links are retained). In this space all nodes represent
routes and the neighbors of any route node are those routes with
which it shares a common station, see Figs. \ref{fig2}{\bf e},
\ref{fig2}{\bf f}.

\begin{table*}[th]
\begin{center}
\begin{tabular}{lrrrrrrrrrrrrrrrrrr}
\hline\hline City &   $\langle k_{\mathds{L}} \rangle$    &
$z_{\mathds{L}}$ & $\ell_{\mathds{L}}^{\rm max}$
         & $\langle \ell_{\mathds{L}} \rangle$ & $\langle {\cal C_{\mathds{L}}}^b\rangle$ &
         $c_{\mathds{L}}$
&   $\langle k_{\mathds{P}} \rangle$    &  $z_{\mathds{P}}$ &
$\ell_{\mathds{P}}^{\rm max}$
         & $\langle \ell_{\mathds{P}} \rangle$ & $\langle {\cal C_{\mathds{P}}}^b\rangle$ &
         $c_{\mathds{P}}$
&   $\langle k_{\mathds{C}} \rangle$    &  $z_{\mathds{C}}$ &
$\ell_{\mathds{C}}^{\rm max}$
         & $\langle \ell_{\mathds{C}} \rangle$ & $\langle {\cal C_{\mathds{C}}}^b\rangle$ &
         $c_{\mathds{C}}$
         \\ \hline
Berlin        &  2.58  & 1.96  &  68  &  18.5 &   2.6$\cdot 10^4$  &  52.8  &  56.61& 11.47 & 5  & 2.9& 2.9$\cdot 10^3$  &  41.9 &  27.56& 4.43&    5  &   2.2 &   1.2$\cdot 10^2$  &  4.75  \\
Dallas        &  2.18  & 1.28  & 156  &  52.0 &  1.4$\cdot 10^5$ &  55.0    & 100.58& 11.23 & 8  & 3.2& 5.9$\cdot 10^3$  &  48.6 &  11.09& 3.45&    7  &   2.7 &    9.2$\cdot 10^1$  & 5.34  \\
D\"usseldorf  &  2.57  & 1.96  &  48  &  12.5 &    8.6$\cdot 10^3$  &  24.4 &  59.01& 10.56 & 5  & 2.6& 1.2$\cdot 10^3$  &  19.7 &  32.18& 2.47&    4  &   1.8 &    4.9$\cdot 10^1$  & 2.23  \\
Hamburg       &  2.65  & 1.85  & 156  &  39.7 &  1.4$\cdot 10^5$  &  254.7   &  50.38&  7.96 &11  & 4.7&1.4$\cdot 10^4$  &  132.2  &  17.51& 4.49&   10  &   4.0 &  9.9$\cdot 10^2$  &   28.3  \\
Hong Kong     &  3.59  & 3.24  &  60  &  11.0 &   1.0$\cdot 10^4$  &  60.3  & 125.67& 10.20 & 4  & 2.2& 1.3$\cdot 10^3$  &  11.7 &  98.98& 2.12&    3  &   1.7 &   1.2$\cdot 10^2$  &  2.14  \\
Istanbul      &  2.30  & 1.54  & 131  &  29.7&   5.7$\cdot 10^4$  &  41.0   &  76.88& 10.59 & 6  & 3.1& 4.2$\cdot 10^3$  &  41.5 &  52.81& 3.86&    5  &   2.3 &   2.6$\cdot 10^2$  &  5.00  \\
London        &  2.60  & 1.87 &  107  & 26.5  & 1.4$\cdot 10^5$   &  320.6   &  90.60& 16.97 & 6  & 3.3& 1.2$\cdot 10^4$ & 90.0& 49.91 & 6.80 &    6 &  2.6& 7.4$\cdot 10^2$ & 11.1\\
Los Angeles   &  2.37 & 1.59 &  210 & 37.1& 7.9$\cdot 10^5$ & 645.3& 97.99 &17.21 &   11 &  4.4& 7.4$\cdot 10^4$ & 399.6& 40.11 & 8.42 &   10 &  3.6& 2.3$\cdot 10^3$ & 22.1\\
Moscow        &  3.32  & 6.25  &  27  &   7.0&   1.1$\cdot 10^4$  &  127.4   &  65.47& 26.48 & 5  & 2.5& 2.7$\cdot 10^3$ &  38.0  & 109.37& 4.57&    4  &   1.9 &   3.2$\cdot 10^2$  &  3.59  \\
Paris         &  3.73  & 5.32  &  28  &   6.4 &   1.0$\cdot 10^4$  &  78.5  &  50.92& 24.06 & 5  & 2.7& 3.1$\cdot 10^3$  &  59.6 &  39.95& 4.67&    4  &   1.9 &   1.1$\cdot 10^2$  &  2.72  \\
Rome          &  2.95  & 2.02  &  87  &  26.4 &   5.0$\cdot 10^4$  &  163.4  &  69.05& 11.34 & 6  & 3.1& 4.2$\cdot 10^3$  &  41.4 &  59.40& 4.86&    5  &   2.5 &   5.1$\cdot 10^2$ &   7.04  \\
Sa\~o Paolo     &  3.21  & 4.17  &  33  &  10.3 &   3.4$\cdot 10^4$  &  268.0  & 137.46& 19.61 & 5  & 2.7& 6.0$\cdot 10^3$  &  38.2 & 151.72& 4.25&    4  &   2.0 &   5.2$\cdot 10^2$  &  4.27  \\
Sydney        &  3.33  & 2.54  &  34  &  12.3 &    7.3$\cdot 10^3$  &  82.9 &  42.88&  7.79 & 7  & 3.0& 1.3$\cdot 10^3$  &  33.6 &  65.02& 2.92&    6  &   2.4 &   3.5$\cdot 10^2$  &  6.30  \\
Taipei        &  3.12  & 2.42  &  74  &  20.9&   5.3$\cdot 10^4$  &  186.2   & 236.65& 12.96 & 6  & 2.4& 3.6$\cdot 10^3$  &  15.4 &  93.33& 2.95&    5  &   1.8 &   1.6$\cdot 10^2$  &  2.44  \\
\hline \hline
\end{tabular}
\end{center}
\caption{PTN characteristics in different spaces (subscripts refer
to $\mathds{L}$, $\mathds{P}$, and $\mathds{C}$-spaces,
correspondingly). $k$:  node degree (nearest neighbors number
$z^{(1)}$); $z=\langle z^{(2)} \rangle/ \langle z^{(1)} \rangle$
($z^{(2)}$ being the next nearest neighbors number); $\ell^{\rm max}$,
$\langle \ell \rangle$: maximal and mean shortest path length (\ref{4.1});
${\cal C}^b$: betweenness centrality (\ref{4.12}); $c$: relation of the mean
clustering coefficient to that of the classical random graph of
equal size (\ref{3.6a}). Averaging has been performed with respect to corresponding network,
only the mean shortest path $\langle \ell \rangle$ is calculated with respect to the
largest connected component. \label{tab2}}
\end{table*}


Below, we will study different features of the PT networks as they
appear when represented in the above defined spaces. It is worthwhile
to mention here, that standard network characteristics being
represented in different spaces turn out to be natural
characteristics one is interested in when judging about the public
transport of a given city. To give an example, the average length of
a shortest path $\langle \ell \rangle$ in $\mathds{L}$-space,
$\langle \ell_{\mathds{L}} \rangle$ gives the number of stops one
has to pass on average to travel between any two stations. When
represented in $\mathds{P}$ space, $\langle \ell_{\mathds{P}}
\rangle$ tells about how many changes one has to do to travel
between any two stations. And, finally, $\langle \ell_{\mathds{C}}
\rangle$ brings about the number of changes one has to do to pass
between any two routes. Another example is given by the node degree
$k$: $k_{\mathds{L}'}$ tells to how many directions a passenger can
travel at a given station; $k_{\mathds{L}}$ is the number of stops
in the direct neighborhood; $k_{\mathds{P}}$ is the number of other
stations reachable without changing a line; whereas
$k_{\mathds{C}}$ tells how many routes are directly accessible from
the given one.

Table \ref{tab2} lists some of the PTN characteristics we
obtained for the cities under consideration using publicly available
data from the web pages of local transport organizations
\cite{database1,database2}. A detailed analysis and discussion
is given in the following sections \ref{III} - \ref{V}.

\section{Local network characteristics} \label{III}

Let us first examine local properties of the PTNs under discussion.
Instead of looking for characteristics of individual nodes
we will be interested in their mean values and statistical
distributions. This approach allows us to derive
conclusions that are significant for the global behavior of the given network.
The simplest but highly important properties are those concerning the node
degrees of a network and in particular their distribution. Early
attempts to model complex networks were performed by mathematicians
 using the concept of random networks \cite{Erdos,Bollobas85} in
which correlations are absent. A wealth of insight was gained by
elaborating the theory on rigorous grounds developing many concepts
which remain among the core of network analysis. A random graph is
given by a set of $N$ nodes and $M$ links. The nodes to which the
two ends of each link are connected are chosen with constant
probability $2M/N$. In case that multiple links are excluded the
average number of neighbors $z_1$ is equal to the average node
degree $k$ which is:
\begin{equation}\label{3.1}
\langle z_1 \rangle = \langle k \rangle = 2M/N.
\end{equation}
 For the node degree $k$ and
its moments $k^m$ the average (\ref{3.1}) can also be considered as
an average with respect to the node degree distribution $p(k)$:
\begin{equation}\label{3.2}
\langle k^m \rangle = \sum_{k=1}^{\rm k^{\rm max}} p(k)  k^m,
\end{equation}
with the obvious notation $k^{\rm max}$ for the maximal node degree.
In (\ref{3.2}), $\langle \dots \rangle$ stands for an ensemble
average over different network configurations. In the following
analyzing empirical data we will often use the same notation for an
average over a large  network instance.
For classical random graphs of finite size the node degree
distribution $p(k)$ is binomial, in the infinite case it becomes a
Poisson distribution.


\begin{figure*}
\centerline{
\includegraphics[width=60mm]{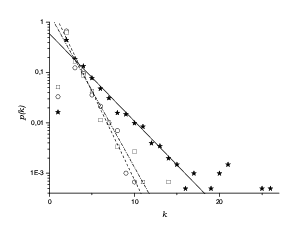}
\includegraphics[width=60mm]{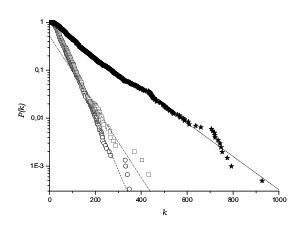}
\includegraphics[width=60mm]{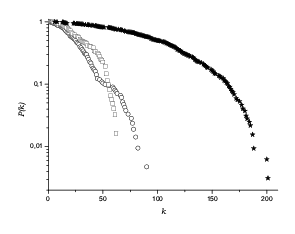}
}
\centerline{{\bf a} \hspace{17em} {\bf b}  \hspace{17em} {\bf c}}
 \caption{{\bf a}: Node degree distributions of PTN of several
 cities in $\mathds{L}$-space. {\bf b}: Cumulative node degree
 distribution in $\mathds{P}$-space. {\bf c}: Cumulative node degree
 distribution in   $\mathds{C}$-space. Berlin (circles, $\hat{k}_{\mathds{L}}=1.24$, $\hat{k}_{\mathds{P}}=39.7$),
 D\"usseldorf (squares, $\hat{k}_{\mathds{L}}=1.43$, $\hat{k}_{\mathds{P}}=58.8$), Hong Kong
(stars, $\hat{k}_{\mathds{L}}=2.50$, $\hat{k}_{\mathds{P}}=125.1$).}
 \label{fig4}
\end{figure*}

\begin{figure*}
\centerline{
\includegraphics[width=60mm]{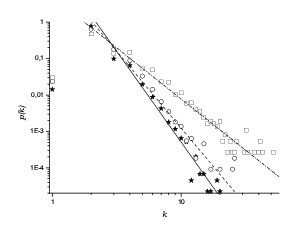}
\includegraphics[width=60mm]{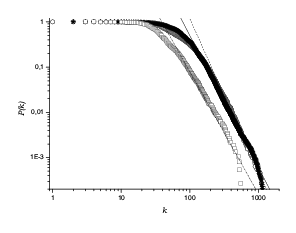}
\includegraphics[width=60mm]{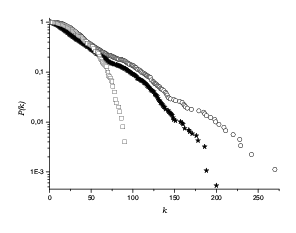}}
\centerline{{\bf a} \hspace{17em} {\bf b}  \hspace{17em} {\bf c}}
 \caption{ {\bf a}: Node degree distributions of PTN of several
 cities in $\mathds{L}$-space.  {\bf b}: Cumulative node degree
 distributions in $\mathds{P}$-space. {\bf c}: Cumulative node degree
 distribution in   $\mathds{C}$-space. London (circles, $\gamma_{\mathds{L}}=4.48$, $\gamma_{\mathds{P}}=4.39$),
 Los Angeles (stars, $\gamma_{\mathds{L}}=4.85$, $\gamma_{\mathds{P}}=3.92$), Paris (squares, $\gamma_{\mathds{L}}=2.62$,
 $\gamma_{\mathds{P}}=3.70$). }
 \label{fig5}
\end{figure*}


In Figs. \ref{fig4}, \ref{fig5} we show node degree distributions
for PTNs of several cities in $\mathds{L}$, $\mathds{P}$, and
$\mathds{C}$-spaces. Note that to get smoother curves we plot in the
case of $\mathds{P}$ and $\mathds{C}$-spaces the cumulative
distributions defined as:
\begin{equation}\label{3.2a}
P(k)= \sum_{q=k}^{k^{\rm max}} p(q).
\end{equation}
In Fig. \ref{fig4} the data is shown in a log-linear plot together
with fits (for $\mathds{L}$ and $\mathds{P}$-spaces) to an
exponential decay:
\begin{equation}\label{3.3}
p(k) \sim \exp(-k/\hat{k}),
\end{equation}
where $\hat{k}$ is of the order of the mean node degree. Within the
accuracy of the data both $\mathds{L}$ and $\mathds{P}$-space
distributions for the cities analyzed in Fig. \ref{fig4} are nicely
fitted by an exponential decay. As far as the $\mathds{L}$-space
data is concerned, we find evidence for an exponential decay for
about half of the cities analyzed, while the other part rather
demonstrate a power law decay of the form:
\begin{equation}\label{3.4}
p(k) \sim 1/k^{\gamma}.
\end{equation}

Figs. \ref{fig5}{\bf a}, \ref{fig5}{\bf b} show the corresponding
plots for three other cities on a log-log scale. Numerical
values of the fit parameters $\hat{k}$ and $\gamma$ (\ref{3.3}),
(\ref{3.4}) for different cities are given in Table \ref{tab2a}.
There, bracketed values indicate a less reliable fit. Note that for
${\mathds{L}}$-space the fit was done directly for the node degree
distribution $p(k)$, whereas due to an essential scattering of data
in ${\mathds{P}}$-space the cumulative distribution (\ref{3.2a}) was
fitted and the corresponding values for the fit parameters
$\gamma_{\mathds{P}}$,  $\hat{k}_{\mathds{P}}$ were extracted from
those for the cumulative distributions.

\begin{table}[ht]
\begin{center}
\tabcolsep3.0mm
\begin{tabular}{lllll}
\hline\hline
City &   $\gamma_{\mathds{L}}$    &  $\hat{k}_{\mathds{L}}$ & $\gamma_{\mathds{P}}$
         & $\hat{k}_{\mathds{P}}$
         \\
         \hline
  Berlin       & (4.30) & 1.24 & (5.85) &  39.7 \\
 Dallas       & 5.49 & (0.78) & (4.67) &  76.9 \\
 Du\"sseldorf & (3.76) & 1.43 & (4.62) &  58.8 \\
 Hamburg      & (4.74) & 1.46 & (4.38) &  60.7 \\
 Hong Kong    & (2.99) & 2.50 & (4.40) & 125.1 \\
 Istanbul     & 4.04 & (1.13) & (2.70) &  71.4 \\
 London       & 4.48 & (1.44) & 4.39 & (143.3) \\
 Los Angeles  & 4.85 & (1.52) & 3.92 & (201.0) \\
 Moscow       & (3.22) & 2.15 & (2.91) &  50.0 \\
 Paris        & 2.62 & (3.30) & 3.70 & (100.0) \\
 Rome         & 3.95 & (1.71) & (5.02) &  54.8 \\
 Sa\~o Paolo  & 2.72 & (4.20) & (4.06) & 225.0 \\
 Sydney       & 4.03 & (1.88) & (5.66) &  38.7 \\
 Taipei       & (3.74) & 1.75 & (5.16) & 201.0 \\
\hline\hline
\end{tabular}
\end{center}
\caption{Parameters of the PTN node degree distributions fit to an exponential (\ref{3.3}) and power law
(\ref{3.4}) behavior. Bracketed values indicate less reliable fits. Subscripts refer to
$\mathds{L}$ and  $\mathds{P}$-spaces \cite{database2}. \label{tab2a}}
\end{table}

While the node degree distribution of almost half of the cities in
the $\mathds{L}$-space representation display a power law decay
(\ref{3.4}), this is not the case for the $\mathds{P}$-space.
So far, the analysis of PTNs of smaller cities never showed any
power-law behavior in $\mathds{P}$-space \cite{Sienkiewicz05,Xu07}.
The data for the three cities shown in Fig. \ref{fig5}{\bf b} gives
first evidence of power law behavior of $P(k)$ in the
$\mathds{P}$-space representation. Previous results concerning
node-degree distributions of PTNs in $\mathds{L}$ and
$\mathds{P}$-spaces seemed to indicate that in general the degree
distribution is power-law like in $\mathds{L}$-space and exponential
in $\mathds{P}$-space. This was interpreted \cite{Sienkiewicz05} as
indicating strong correlations in $\mathds{L}$-space and random
connections between the routes explaining $\mathds{P}$-space
behavior. Our present study, which includes a much less homogeneous
selection of cities  (Ref. \cite{Sienkiewicz05} was based on
exclusively Polish cities) shows that almost any combination of
different distributions in $\mathds{L}$ and $\mathds{P}$-spaces may
occur. However, the three cities that show a power law
distribution in $\mathds{P}$-space also exhibit power law behavior
in $\mathds{L}$-space, as one can see comparing Figs. \ref{fig5}{\bf
a} and \ref{fig5}{\bf b}.

In $\mathds{C}$-space the decay of the node degree distribution is
exponential or faster, as one can see from the plots in Fig.
\ref{fig4}c and \ref{fig5}c. From the cities presented
there, only the PTNs of Berlin, London, and Los Angeles are governed
by an exponential decay and their node degree distributions can be
approximated by a straight line in the figures.

For most cities that show a power law degree distribution in
$\mathds{L}$-space the corresponding exponent $\gamma_{\mathds{L}}$
is $\gamma_{\mathds{L}} \sim 4$. Note that the exponents found for
the PTNs of Polish cities of similar size $N$ also lie in this
region: $\gamma_{\mathds{L}} =3.77$ for Krakow (with number of
stations $N=940$), $\gamma_{\mathds{L}} =3.9$ for Lodz ($N=1023$),
$\gamma_{\mathds{L}} =3.44$ for Warsaw ($N=1530$)
\cite{Sienkiewicz05}. According to the general classification of
scale-free networks \cite{Dorogovtsev02} this indicates that in many
respect these networks are expected to behave similar to those with
exponential node degree distribution. Prominent exceptions to this
rule are provided by the PTNs of Paris ($\gamma_{\mathds{L}}=2.62$)
and Sa\~o-Paolo ($\gamma_{\mathds{L}}=2.72$). Furthermore, values of
$\gamma_{\mathds{L}}$ in the range $2.5\div3.0$ were recently
reported for the bus networks of three cities in China: Beijing
($N=3938$),  Shanghai ($N=2063$),
 and Nanjing ($N=1150$) \cite{Xu07}.

\begin{figure}
\centerline{
\includegraphics[width=90mm]{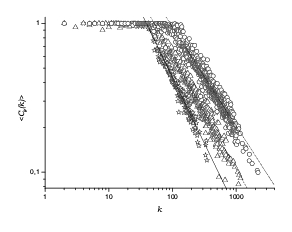}
 }
\caption{Mean clustering coefficient $\langle C_{\mathds{P}}(k)
\rangle$ of several PTN in $\mathds{P}$-space. Berlin
(stars), London (triangles), Taipei (circles).}
 \label{fig6}
\end{figure}

The connectivity within the closest neighborhood of a given node is
described by the clustering coefficient defined as
\begin{equation}\label{3.5}
C_i = \frac{2y_i}{k_i(k_i-1)},
\end{equation}
where $y_i$ is the number of links between the $k_i$ nearest
neighbors of the node $i$.
The clustering coefficient of a node may also be defined as the
probability of any two of its randomly chosen neighbors to be
connected.
For the mean value of the clustering
coefficient of a random graph one
finds
\begin{equation}\label{3.6}
\langle C \rangle^{\cal R} = \frac{\langle k \rangle^{\cal R}}{N}
= \frac{2M}{N^2}.
\end{equation}
In Table \ref{tab2} we give the values of the mean clustering
coefficient  in ${\mathds{L}}$, ${\mathds{P}}$, and
${\mathds{C}}$-spaces. The highest absolute values of the clustering
coefficient are found in ${\mathds{P}}$-space, where their  range is
given by $\langle C_{\mathds{P}} \rangle=0.7\div0.9$ (c.f. with
$\langle C_{\mathds{L}} \rangle=0.02\div0.1$). This is due to the
fact that in this space each route gives rise to a fully connected
subgraph (complete graph). In order to make numbers comparable we
normalize the value of $\langle C \rangle$ by the mean clustering
coefficient (\ref{3.6}) of a random graph of the same size:
\begin{equation}\label{3.6a}
c=N^2\langle C \rangle/(2M).
\end{equation}
In $\mathds{L}$ and $\mathds{P}$-representations we find the mean
clustering coefficient to be larger by orders of magnitude relative
to the random graph. This difference is less pronounced in
${\mathds{C}}$-space indicating a lower degree of organization in
these networks. Furthermore, we find these values to vary strongly
within the sample of the 14 cities. This suggests that the concepts
according to which various PTNs are structured lead to a measurable
difference in their organization.

\begin{figure*}
\centerline{
\includegraphics[width=60mm]{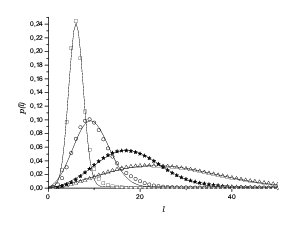}
\includegraphics[width=60mm]{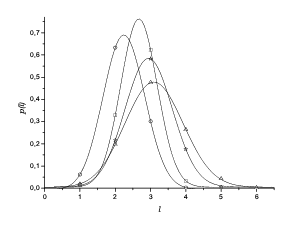}
\includegraphics[width=60mm]{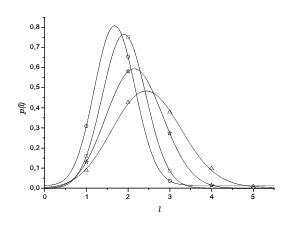}
 }
 \centerline{
{\bf a} \hspace{17em} {\bf b} \hspace{17em} {\bf c}
 }
\caption{Mean shortest path length distribution $P(\ell)$ for the PTN of
Berlin (stars), Hong Kong (circles), Paris (boxes) and Rome (triangles). Solid line
shows a fit to the function (\ref{4.3}). {\bf a}:
$\mathds{L}$-space; {\bf b}: $\mathds{P}$-space;  {\bf c}: $\mathds{C}$-space.}
 \label{fig7}
\end{figure*}

 In $\mathds{P}$-space the clustering coefficient of a
node is strongly correlated with the node degree. In Fig. \ref{fig6}
we show the mean clustering coefficient of nodes of degree $k$,
$\langle C_{\mathds{P}}(k) \rangle$, as a function of $k$ for
several PTNs. Its behavior can be understood as follows. Recall that
 the $\mathds{P}$-space the degree of a node (station) equals
the number of stations that can be reached from a given one. Each
route enters the network as a complete graph, within which every
node has a clustering coefficient of one. A small number $k$ of
neighbors of a given station indicates that the station belongs to a
single route (i.e. $\langle C_{\mathds{P}}(k) \rangle $ is most
probably equal to one). For nodes with higher degrees $k$ it is more
probable that they belong to more than one route. Consequently,
$\langle C_{\mathds{P}}(k) \rangle $ decreases with $k$. The change in
the behavior of $\langle C_{\mathds{P}}(k) \rangle$ should occur at
some value of $k$ which is of the order of the mean number of stops
of the routes. The prominent feature of the function $\langle
C_{\mathds{P}}(k) \rangle$ in $\mathds{P}$-space is that it decays
following a power law
 \begin{equation} \label{3.7}
\langle C_{\mathds{P}}(k) \rangle  \sim~k^{-\beta}.
 \end{equation}
 Within a simple model of networks with star-like topology this
 exponent is found to be of value $\beta=1$ \cite{Sienkiewicz05}.
 In transport networks. This behavior was first observed for the
Indian railway network \cite{Sen03} and then for the Polish PTNs
\cite{Sienkiewicz05}. In our case, the values of the exponent
$\beta$ for the networks studied  lie  in the range from
$0.65$ (Sa\~o Paolo) to $0.96$ (Los Angeles) with a mean value of
$0.82$.

\section{Global characteristics}\label{IV}

\subsection{Path length distribution}\label{IVa}

\begin{figure}
\centerline{
\includegraphics[width=90mm]{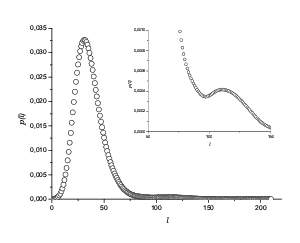}
 }
 \caption{Mean shortest path length distribution in $\mathds{L}$-space,
$P_\mathds{L}(\ell)$, for the PTN  of Los Angeles.}
 \label{fig8}
\end{figure}

Let $\ell_{i,j}$ be the length of a shortest path between sites $i$
and $j$ in a given space. The mean shortest path is defined as
\begin{equation}\label{4.1}
\langle\ell \rangle = \frac{2}{N(N-1)}\sum_{i
> j=1}^N \ell_{ij}.
\end{equation}
Note that $\langle\ell \rangle$ is well-defined only if nodes $i$
and $j$ belong to the same connected component of the network. In the
following any expression as given in Eq. (\ref{4.1}) will be
restricted to this case. Furthermore, related network
characteristics will be calculated for the largest (or giant)
connected component, GCC. Correspondingly, $N$ denotes the number of
constituting nodes of this component. Denoting the path length distribution as
$P(\ell)$, the average (\ref{4.1}) reads
\begin{equation}\label{4.2}
\langle\ell \rangle = \sum_{\ell=1}^{\ell^{\rm max}} P(\ell)\ell,
\end{equation}
where $\ell^{\rm max}$ is maximal shortest path length found on the
connected component. In Fig. \ref{fig7} we plot the mean shortest
path length distributions obtained in different spaces for several
selected cities. Together with the data we plot a fit to the
asymmetric unimodal distribution \cite{Sienkiewicz05}:
\begin{equation}\label{4.3}
P(\ell)=A \ell \exp{(-B\ell^2+C\ell)},
\end{equation}
where $A,B,C$ are fit parameters. As can be seen from the
figures, the data is generally nicely reproduced by this ansatz.
However, in certain networks additional features may lead to a
deviation from this behavior as can be seen from Fig. \ref{fig8},
which shows the mean shortest path length distribution in
$\mathds{L}$-space $P_{\mathds{L}}(\ell)$ for Los Angeles. One
observes a second local maximum on the right shoulder of the
distribution. Qualitatively this behavior may be explained by
assuming that the PTN consists of more than one community. For the
simple case of one large community and a second smaller one at some
distance this situation will result in short intra-community paths
which will give rise to a global maximum and a set of longer paths
that connect the larger to the smaller community resulting in
additional local maxima. Such a situation definitely appears to be
present in the case of the Los Angeles PTN, see Fig. \ref{fig1}.

\begin{figure}
\centerline{
\includegraphics[width=45mm]{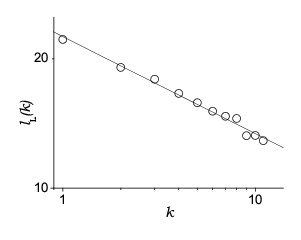}
\includegraphics[width=45mm]{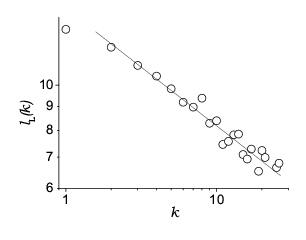}
 }
\centerline{{\bf a} \hspace{13em} {\bf b}}
\centerline{\includegraphics[width=45mm]{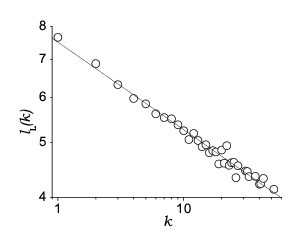}
\includegraphics[width=45mm]{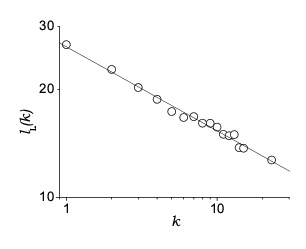}
 }
 \centerline{{\bf c} \hspace{13em} {\bf d}}
\caption{Mean path $\ell_{\mathds{L}}(k)$ (\ref{4.6}) in the
${\mathds{L}}$-space  as a function of the node degree $k$ with a fit to the power
law decay (\ref{4.6a}). {\bf a}: Berlin, $\alpha_{\mathds{L}}=0.23$;
{\bf b}: Hong Kong, $\alpha_{\mathds{L}}=0.25$; {\bf c}: Paris, $\alpha_{\mathds{L}}=0.15$;
{\bf d}: Taipei, $\alpha_{\mathds{L}}=0.23$.}
 \label{fig9}
\end{figure}

Let us introduce a characteristic that informs how remote a
given node is from the other nodes of the networks. For the node $i$ this
may be characterized by the value:
\begin{equation}\label{4.4}
\ell_i = \frac{1}{N-1}\sum_{j\neq i} \ell_{ij}.
\end{equation}
Now, the mean shortest path (\ref{4.1}) can be defined in terms of
$\ell_i$ as:
\begin{equation}\label{4.5}
\langle\ell \rangle = \frac{1}{N}\sum_{i} \ell_{i}.
\end{equation}
In order to look for correlations between $\ell_i$ and  the node
degree $k_i$ let us introduce the value:
\begin{equation}\label{4.6}
 \ell (k)  = \frac{1}{N_k}\sum_{i=1}^{N} \ell_i
\delta_{k,k_i},
\end{equation}
where $N_k$ is number of nodes of degree $k$ and $\delta_{k,k_i}$ is the
Kronecker delta. Consequently, $\ell (k)$ is the mean shortest path
length between any node of degree $k$ and other nodes of the
network. For the majority of the analyzed cities the dependence of the
mean path $\ell_{\mathds{L}}(k)$ (\ref{4.6}) on the node degree $k$
in $\mathds{L}$-space can be approximated by a power law
\begin{equation}\label{4.6a}
\ell_{\mathds{L}}(k)\sim k^{-\alpha_{\mathds{L}}}.
\end{equation}
 The value of
the exponent varies in the range $\alpha_{\mathds{L}}=0.17\div
0.27$. We show this dependence for several cities in Fig.
\ref{fig9}.

\begin{figure}[h]
\centerline{
\includegraphics[width=70mm]{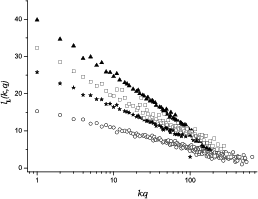}
}
 \caption{Mean path $\ell_{\mathds{L}}(k,q)$ (\ref{4.7}) in the $\mathds{L}$-space
 as a function of $kq$ for the PTN
of Berlin (stars), Hong Kong (circles), Rome (triangles), and Taipei
(squares).}
 \label{fig10}
\end{figure}

A particular relation between
path lengths and node degrees can be shown to hold relating the mean path
length between two nodes to the product of their node
degrees. To this end let us define
\begin{equation}\label{4.7}
 \ell (k,q)  = \sum_{i,j=1}^{N} \ell_{i,j}
\delta_{k_ik_j,kq}.
\end{equation}
As has been shown in \cite{Holyst05}, this relation can be
approximated by
\begin{equation}\label{4.8}
 \ell (k,q)  = A - B \log (kq).
\end{equation}
For random networks the coefficients $A$ and $B$ can be calculated
exactly \cite{Fronczak03}. The validity of Eq. (\ref{4.8}) was
checked on the base of PTNs of some Polish cities and a rather good
agreement for the majority of the cities was found in
${\mathds{L}}$-space. In our analysis which concerns PTNs of much
larger size, we do not observe the same good agreement for all
cities. The suggested logarithmic dependence (\ref{4.8}) was found
by us in ${\mathds{L}}$-space also for the larger cities, however
with much more pronounced scatter of data for large values of the
product $kq$. In Fig. \ref{fig10} we plot the mean path
$\ell_{\mathds{L}}(k,q)$  in the ${\mathds{L}}$-space for the PTN of
Berlin, Hong Kong, Rome, and Taipei. Note, however, that due to the
scatter of data a logarithmic dependence  frequently is
indistinguishable from a power law with a small exponent.

\begin{figure}
\centerline{
\includegraphics[width=45mm]{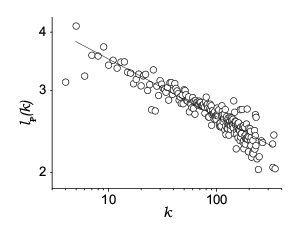}
\includegraphics[width=45mm]{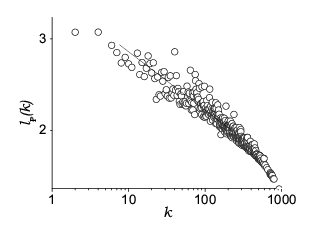}
 }
 \centerline{{\bf a} \hspace{13em} {\bf b}}
\centerline{\includegraphics[width=45mm]{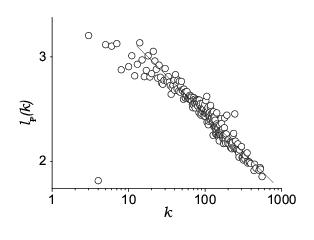}
\includegraphics[width=45mm]{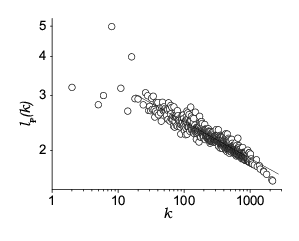}
 }
 \centerline{{\bf c} \hspace{13em} {\bf d}}
\caption{Mean path $\ell_{\mathds{P}}(k)$ in $\mathds{P}$-space as a
function of the node degree $k$ and its fit to the power law decay (\ref{4.8a}).
{\bf a}: Berlin, $\alpha_{\mathds{P}}=0.13$;
{\bf b}: Hong Kong, $\alpha_{\mathds{P}}=0.12$; {\bf c}: Paris, $\alpha_{\mathds{P}}=0.13$;
{\bf d}: Taipei, $\alpha_{\mathds{L}}=0.12$.
}
 \label{fig11}
\end{figure}

\begin{figure}
\centerline{
\includegraphics[width=45mm]{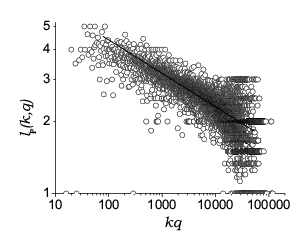}
\includegraphics[width=45mm]{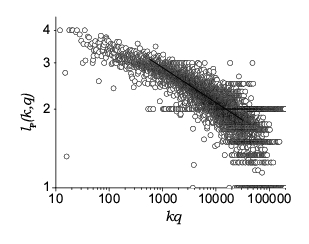}
}
 \centerline{{\bf a} \hspace{13em} {\bf b}}
\caption{Mean path $\ell_{\mathds{P}}(k,q)$ in $\mathds{P}$-space
for PTN of Berlin ({\bf a}) and Paris ({\bf b}) as a function of $kq$.}
 \label{fig12}
\end{figure}

In $\mathds{P}$-space, the shortest path length $\ell_{ij}$ gives
the minimal  number of routes required to be used in order to reach
site $j$ starting from the site $i$. In turn, $\ell_i$, Eq.
(\ref{4.4}), defines the number of routes one uses on average
traveling from the site $i$ to any  node of the network. The higher
the node degree, the easier it is to access  other routes in the
network. Therefore, also in $\mathds{P}$-space one expects a
decrease of $\ell_{\mathds{P}}(k)$ when $k$ increases. This is shown
for several cities in Fig. \ref{fig11}. Besides the expected
decrease of $\ell_{\mathds{P}} (k)$, one can see a tendency to a
power-law decay
\begin{equation}\label{4.8a}
\ell_{\mathds{P}} (k) \sim k^{-\alpha_{\mathds{P}}}.
\end{equation}
 The value
of the exponent $\alpha_{\mathds{P}}$ varies in the interval
$\alpha_{\mathds{P}}=0.09$ (for Sydney) to
$\alpha_{\mathds{P}}=0.17$ (for Dallas) and is centered around
$\alpha_{\mathds{P}}=0.12\div 0.13$ as shown for the cities in Fig.
\ref{fig11}. The mean path $\ell_{\mathds{P}}(k,q)$ as a function of
$kq$ for several cities is given in ${\mathds{P}}$-space in Fig.
\ref{fig12}. The scattering of data is much more pronounced than in
$\mathds{L}$-space. However one distinguishes a tendency of
$\ell_{\mathds{P}}(k,q)$ to decrease with an increase of $kq$. The
red lines in Figs. \ref{fig12} are the guides to the eye
characterizing the decay.

\subsection{Centralities}\label{IVb}

To measure the importance of a given node with respect to different
properties of a graph a number of so-called centrality measures have
been introduced
\cite{Brandes01,Sabidussi66,Hage95,Shimbel53,Freeman77}. Most of
them are based on either measuring path lengths to other nodes or on
counting the number of paths between other nodes mediated by this
node.
The closeness ${\cal C}^c(i)$ \cite{Sabidussi66} and graph
${\cal C}^g(i)$ \cite{Hage95} centralities of a node $i$ are based
on the shortest path lengths $\ell_{ij}$ to other nodes $j$:
\begin{eqnarray}\label{4.9}
{\cal C}^c(i)&=&\frac{1}{\sum_{j\neq i} \ell_{i,j}}, \\ \label{4.10}
 {\cal C}^g(i)&=&\frac{1}{{\rm
max}_{j\neq i} \ell_{i,j}}.
 \end{eqnarray}
 Only nodes $j$ that belong to the same connected component as $i$
 contribute to (\ref{4.9}), (\ref{4.10}).
For a given node these properties obviously depend on the size of
the connected component to which the node belongs.
 The importance of the node
 $i$ with respect to the connectivity within the graph may be
 measured in terms of the number of shortest paths $\sigma_{jk}(i)$
 between nodes $j$ and $k$ that go via node $i$. Denoting by $\sigma_{jk}$
 the overall number of shortest paths between nodes $j$ and $k$ one
 defines stress ${\cal C}^s(i)$ \cite{Shimbel53} and betweenness ${\cal C}^b(i)$
\cite{Freeman77} centralities by:
\begin{eqnarray}\label{4.11}
 {\cal C}^s(i)&=& \sum_{j\neq i \neq k} \sigma_{jk}(i), \\ \label{4.12}
 {\cal C}^b(i)&=&\sum_{j\neq i \neq k} \frac{\sigma_{jk}(i)}{\sigma_{jk}}.
 \end{eqnarray}
Numerical values of the betweenness centrality (\ref{4.12}) are given
in Table \ref{tab1} in $\mathds{L}$, $\mathds{P}$ and
$\mathds{C}$-spaces.

\begin{figure*}
\centerline{
\includegraphics[width=55mm]{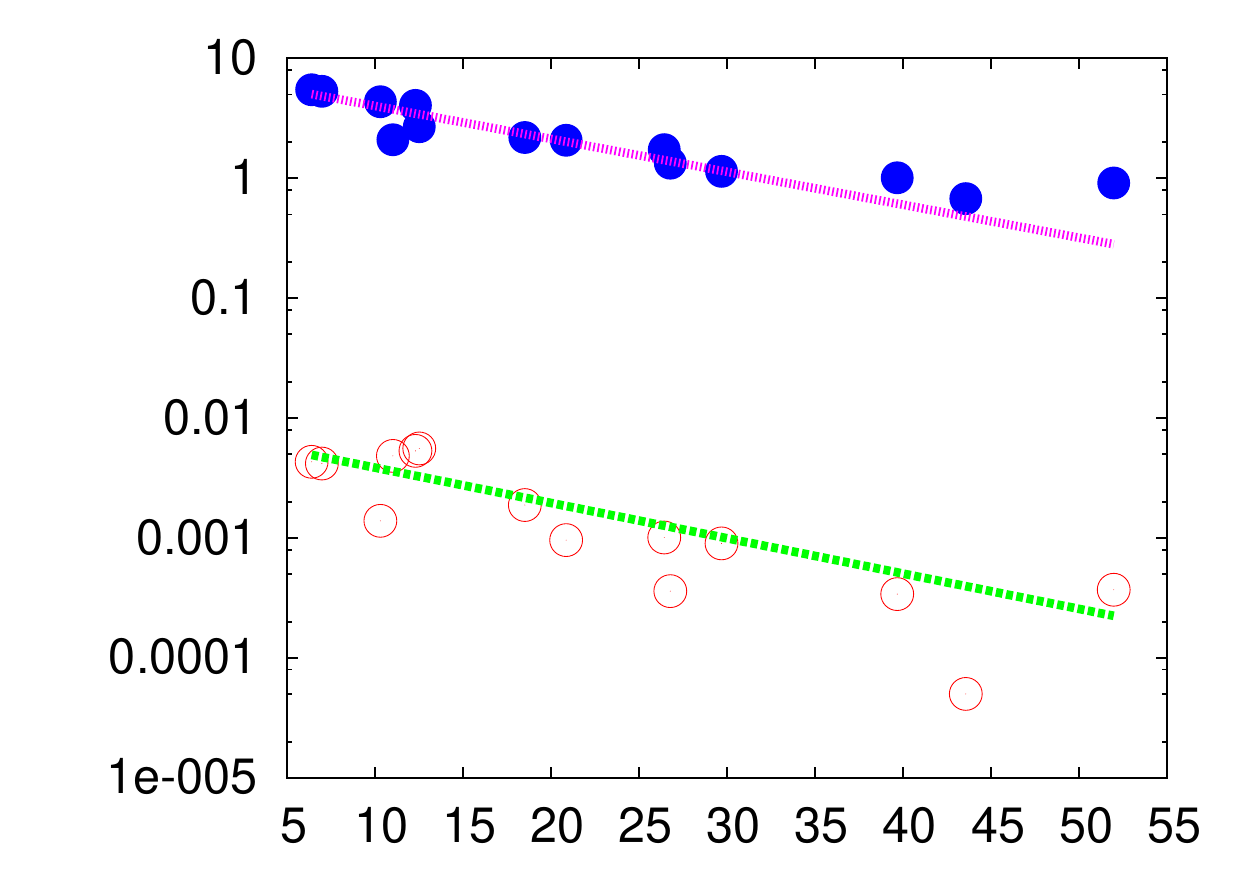}
\includegraphics[width=55mm]{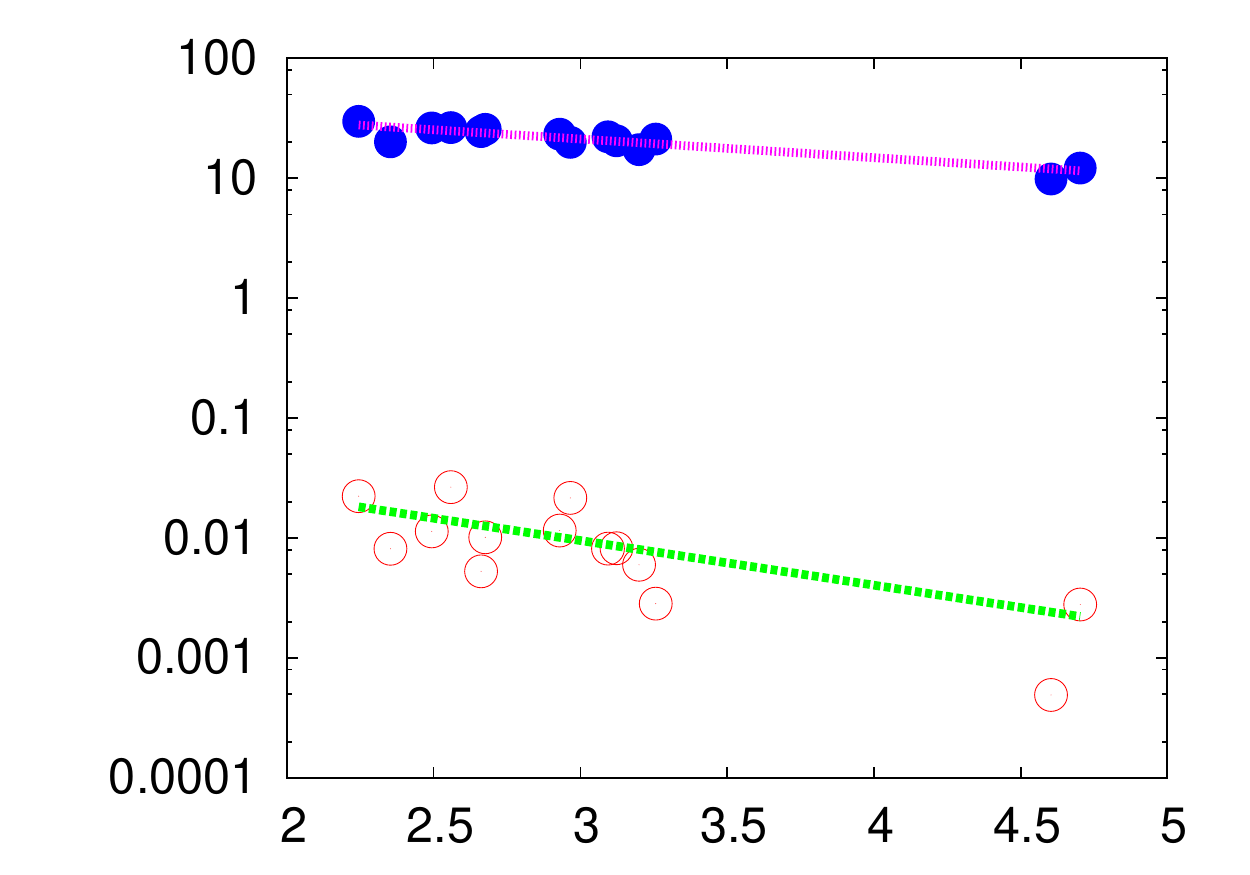}
\includegraphics[width=55mm]{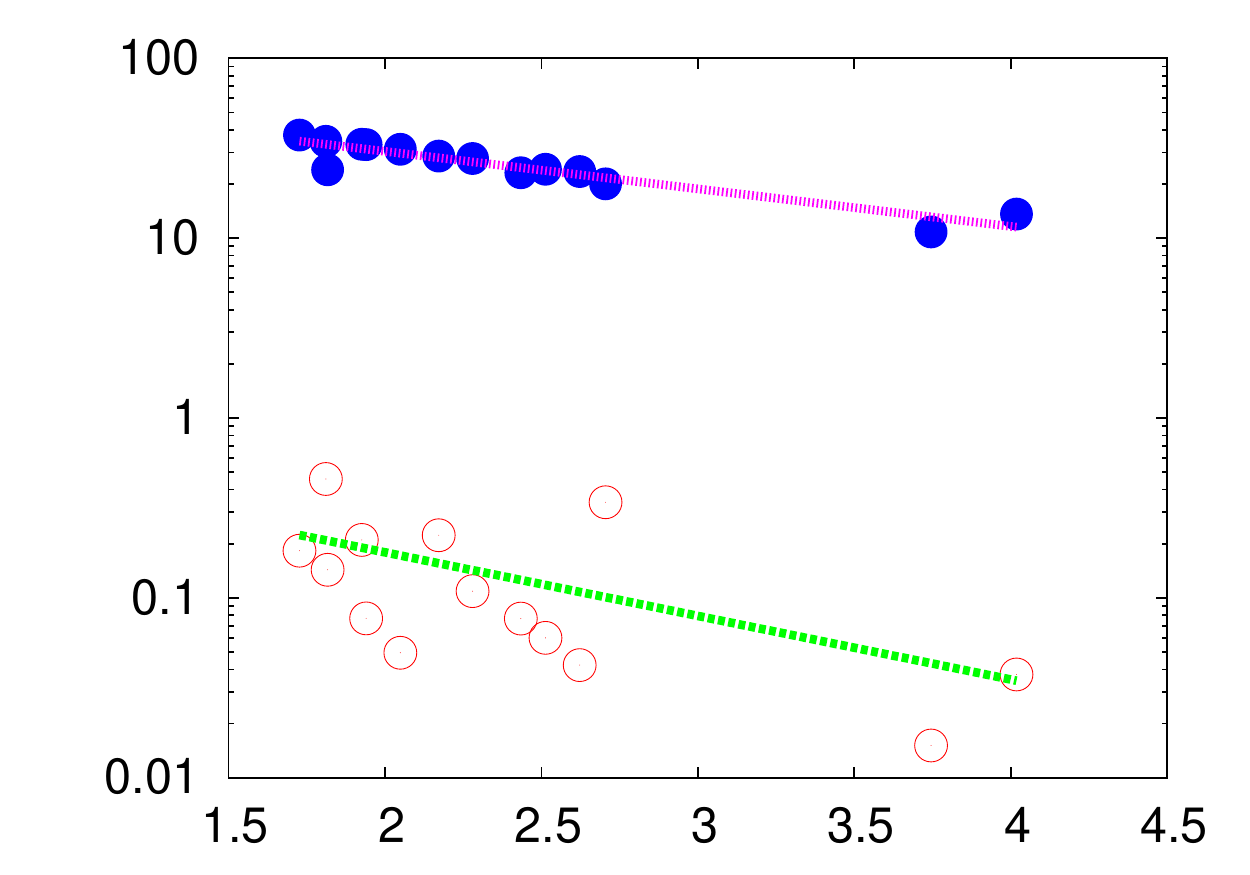}
}
\centerline{
{\bf a} \hspace{17em} {\bf b} \hspace{17em} {\bf c}
 }
 \caption{Correspondence between the mean shortest path $\langle \ell \rangle$
and mean centralities $\langle {\cal C}^c \rangle$ (open circles), $\langle {\cal C}^g \rangle$
(filled circles) for all fourteen PTN listed in Table \ref{tab2} in
({\bf a}) $\mathds{L}$, ({\bf b}) $\mathds{P}$, and ({\bf c}) $\mathds{C}$-spaces.}
 \label{fig13}
\end{figure*}

Averaging the two centralities that are based on path length
(\ref{4.9}), (\ref{4.10}) one obtains values that are closely
related to the average shortest path length on the GCC. As far as
this relation is independent of the representation  of the PTN, we
find very similar correspondence between $\langle \ell \rangle$ and
the mean centralities $\langle {\cal C}^c \rangle$, $\langle {\cal
C}^g \rangle$ in all spaces considered as shown in  Fig.
\ref{fig13}. The fact that these centralities are based on the
inverse path length is reflected by the negative slope of the curves
shown in the figures.


\begin{figure}
\centerline{
\includegraphics[width=45mm]{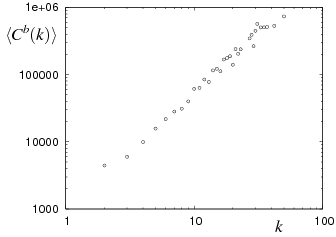}
\includegraphics[width=45mm]{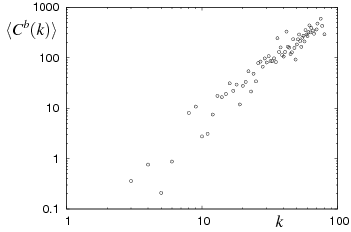}}
\centerline{ {\bf a} \hspace{15em} {\bf b}
 }
\centerline{
\includegraphics[width=45mm]{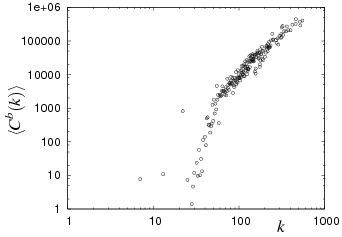}
\includegraphics[width=45mm]{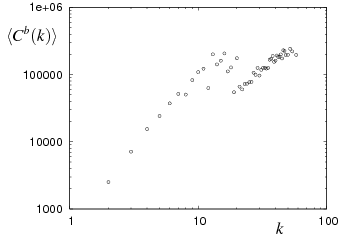}}
\centerline{ {\bf c} \hspace{15em} {\bf d}
 }
\caption{Mean betweenness centrality $\langle {\cal C}^b(k)\rangle$ -
degree $k$ correlations for the PTN of Paris in  ({\bf a}) $\mathds{L}$,
({\bf b}) $\mathds{C}$, ({\bf c}) $\mathds{P}$, and ({\bf d}) $\mathds{B}$-spaces.}
 \label{fig14}
\end{figure}

\begin{figure}
\centerline{
\includegraphics[width=48mm]{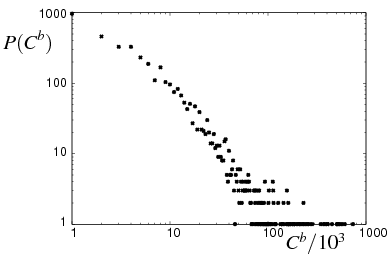}
 }
\caption{Betweenness centrality ${\cal C}^b$ - distribution for PTN of Paris in $\mathds{L}$ - space.}
 \label{fig14a}
\end{figure}

The betweenness centrality (\ref{4.12}) and the related stress
centrality (\ref{4.11}) of a given node measure the share of the
mean paths between nodes that are mediated by that node. It is
obvious that a node with a high degree has a higher probability to
be part of any path connecting other nodes. This relation between
${\cal C}^b$ and the node degree may be quantified by observing
their correlation. In Figs. \ref{fig14} we plot the mean betweenness
centrality $\langle {\cal C}^b(k)\rangle$ of all nodes that have a
given degree $k$. There, we present results for the PTN of Paris in
$\mathds{L}$, $\mathds{C}$ and $\mathds{P}$, and
$\mathds{B}$-spaces. Especially well expressed is the
betweenness-degree correlation in $\mathds{L}$-space (Fig.
\ref{fig14}{\bf a}) and with somewhat less precision in
$\mathds{C}$-space (Fig. \ref{fig14}{\bf b}). In both cases there is
a clear tendency to a power law $\langle {\cal C}^b(k)\rangle \sim
k^\eta$ with an exponent $\eta=2\div 3$.
Let us note here, that this power law together with the scale free
behavior of the degree distribution implies that also the betweenness
distribution should follow a power law with an exponent $\delta$.
This behavior is clearly identified in Fig.\ref{fig14a} for the
$\mathds{L}$-space betweenness distribution of the Paris PTN, for which we
find an exponent $\delta\approx 1.5$. The resulting scaling relation
\cite{Goh01}
\begin{equation}
\eta = (\gamma-1)/(\delta-1)
\end{equation}
is fulfilled within the accuracy for these exponents.
In the plots for both
$\mathds{B}$ and $\mathds{P}$-spaces we observe the occurrence  of
two regimes which correspond to small and large degrees $k$. This
separation however has a different origin in each of these cases. In
the $\mathds{B}$-space representation, the network consists of nodes
of two types, route nodes and station nodes. Typically, station
nodes are connected only to a low number of routes while there is a
minimal number of stations per route. One may thus identify the low
degree behavior as describing the betweenness of station nodes,
while the high degree behavior corresponds to that of route nodes.
In the overlap region of the two regimes one may observe that when
having the same degree station nodes have a higher betweenness than
route nodes. The occurrence of two regimes in the $\mathds{P}$-space
representation has a similar origin as the change of behavior
observed for the mean clustering coefficient $\langle
C_{\mathds{P}}(k) \rangle$, see Fig.\ref{fig6}. Namely, stations
with low degrees in general belong only to a single route and thus
are of low importance for the connectivity within the network
resulting in a low betweenness centrality. Comparing our results
with those of Ref. \cite{Sienkiewicz05} we do not however find a
saturation for the low $k$ region, as observed there. Similar
betweenness $\langle {\cal C}^b(k)\rangle$ - degree relations as
observed in Fig. \ref{fig14} for the PTN of Paris we also find for
most of the other cities, however, with different quality of expression.

\subsection{Harness} \label{IVc}

\begin{figure}
\centerline{
\includegraphics[width=45mm]{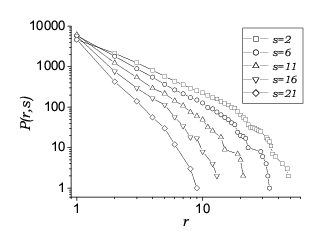}
\includegraphics[width=45mm]{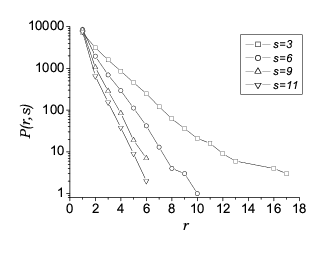}}
\centerline{{\bf a} \hspace{17em} \bf{b} }
 \caption{Cumulative harness distributions for Istanbul ({\bf a}) and for Moscow ({\bf b}) PTN.}
 \label{fig15}
\end{figure}

\begin{figure}
\centerline{
\includegraphics[width=45mm]{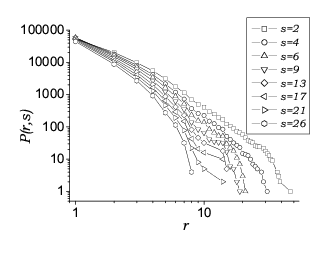}
\includegraphics[width=45mm]{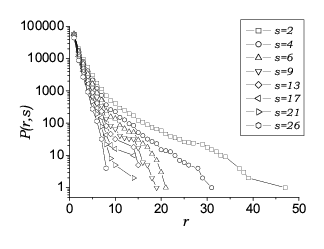}}
\centerline{{\bf a} \hspace{15em} {\bf b}}
 \caption{Cumulative harness distributions for Los Angeles. {\bf a}: log-log scale; {\bf b}: log-linear scale.}
 \label{fig16}
\end{figure}

Besides the local and global properties of networks described
above which can be defined in any type of network, there are some
characteristics that are unique for PTNs and networks with similar
construction principles. A particularly striking example is the fact
that as far as the routes share the same grid of streets and tracks
often a number of routes will proceed in parallel along shorter or
longer sequences of stations. Similar phenomena are observed in
networks built with real space consuming links such as cables,
pipes, neurons, etc. In the present case this behavior may be easily
worked out on the basis of sequences of stations serviced by each
route. To quantify this behavior recently the notion of network
harness has been introduced \cite{Ferber07a}. It is described by the
harness distribution $P(r,s)$: the number of sequences of $s$
consecutive stations that are serviced by $r$ parallel routes.
Similarly to the node-degree distributions, we observe that the
harness distribution for some cities (Hong Kong, Istanbul, Paris,
Rome, Sa\~o Paolo, Sydney) may be fitted by a power law:
 \begin{equation} \label{4.12a}
 P(r,s) \sim r^{-\gamma_s}, \hspace{2em} \mbox{for fixed $s$},
 \end{equation}
 whereas the PTNs of other cities (Berlin, Dallas, D\"usseldorf, London, Moscow)
are better fitted to an exponential decay:
 \begin{equation} \label{4.13}
 P(r,s) \sim \exp{(-r/\hat{r}_s)}, \hspace{2em} \mbox{for fixed $s$}.
 \end{equation}
 As examples we show the harness distribution for Istanbul (Fig.
\ref{fig15}{\bf a}) and for Moscow (Fig. \ref{fig15}{\bf b}).
Moreover, sometimes (we observe this for Los Angeles and Taipei),
for larger $s$ the regime (\ref{4.12a}) changes to (\ref{4.13}). We
show this for the PTN of Los Angeles in Fig. \ref{fig16}. There, one
can see that for small values of $s$ the curves are better fitted to
a  power law dependence (\ref{4.12a}). With increasing $s$ a
tendency to an exponential decay (\ref{4.13}) appears: Fig.
\ref{fig16}{\bf b}.

As one can observe from the Figs. \ref{fig15}, \ref{fig16} the slope
of the harness distribution $P(r,s)$ as a function of the number of
routes $r$ increases with an increase of the sequence length $s$.
For PTNs for which the harness distribution follows power law
(\ref{4.12a}) the corresponding exponents $\gamma_s$ are found in
the range of $\gamma_s=2\div 4$. For those distributions with an
exponential decay the scale $\hat{r}_s$ (\ref{4.13}) varies in the
range $\hat{r}_s=1.5\div 4$. The power laws observed for the
behavior of $P(r,s)$ indicate a certain level of organization and
planning which may be driven by the need to minimize the costs of
infrastructure and secondly by the fact that points of interest tend
to be clustered in certain locations of a city. Note that this
effect may be seen as a result of the strong interdependence of the
evolutions of both the city and its PTN.

We want to emphasize that the harness effect is a feature of the
network given in terms of its routes but it is invisible in any of
the graph representations presented so far. In particular PTN
representation in terms of a simple graph which do not contain
multiple links (such as $\mathds{L}$, $\mathds{P}$, $\mathds{C}$ and
$\mathds{B}$-spaces) can not be used to extract harness behavior.
Furthermore, the multi-graph representations (such as $\mathds{L}'$,
$\mathds{P}'$, and $\mathds{C}'$-spaces) would need to be extended
to account for the continuity of routes. As noted above, the notion
of harness may be useful also for the description of other networks
with similar properties. On the one hand, the harness distribution
is closely related to distributions of flow and load on the network.
On the other hand, in the situation of space-consuming links (such
as tracks, cables, neurons, pipes, vessels) the information about
the harness behavior may be important with respect to the spatial
optimization of networks.

A generalization may be readily formulated to account for real-world
networks in which links (such as cables) are organized in parallel
over a certain spatial distance. While for the PTN this distance is
simply measured by the length of a sequence of stations, a more
general measure would be the length of the contour along which these
links proceed in parallel.

\section{Generalized assortativities}\label{V}

To describe correlations between the properties of neighboring nodes
in a network the notion of assortativity was introduced measuring
the correlation between the node degrees of neighboring nodes in
terms of the mean Pearson correlation coefficient
\cite{Newman02,Newman03b}. Here, we propose to generalize this
concept to also measure correlations between the values of other
node characteristics (other observables). For any link $i$ let $X_i$
and $Y_i$ be the values of the observable at the two nodes connected
by this link. Then the correlation coefficient is given by:
\begin{equation} \label{5.1}
r = \frac{M^{-1}\sum_i X_iY_i - [M^{-1}\sum_i \frac{1}{2}(X_i+Y_i)]^2}%
{M^{-1}\sum_i \frac{1}{2}(X_i^2+Y_i^2)-[M^{-1}\sum_i \frac{1}{2}(X_i+Y_i)]^2}
\end{equation}
where summation is performed with respect to the $M$ links of the
network. Taking $X_i$ and $Y_i$ to be the node degrees Eq.
(\ref{5.1}) is equivalent to the usual formula for the assortativity
of a network \cite{Newman02}. Here we will call this special case
the degree assortativity $r^{(1)}$. In the following we will
investigate correlations between other network characteristics such
as the observables considered above, $z_2$, $C_i$ (\ref{3.5}), ${\cal
C}^c$ (\ref{4.9}), ${\cal C}^g$ (\ref{4.10}), ${\cal C}^s$
(\ref{4.11}), ${\cal C}^b$ (\ref{4.12}). Consequently, this results
in generalized assortativities of next nearest neighbors
($r^{(2)}$), clustering coefficients ($r^{cl}$), closeness ($r^c$),
graph ($r^g$), stress ($r^s$), and betweenness ($r^b$) centralities.

The numerical values of the above introduced assortativities
$r^{(1)}$ and $r^{(2)}$ for the PTN under discussion are listed in
Table \ref{tab3} in $\mathds{L}$, $\mathds{P}$ and
$\mathds{C}$-spaces. With respect to the values of the standard node
degree assortativity $r^{(1)}_{\mathds{L}}$ in $\mathds{L}$-space,
we find two groups of cities. The first is characterized by values
$r^{(1)}_{\mathds{L}}=0.1\div 0.3$. Although these values are still
small they signal a finite preference for assortative mixing. That
is, links tend to connect nodes of similar degree. In the second
group of cities these values are very small
$r^{(1)}_{\mathds{L}}=-0.02 \div 0.08$ showing no preference in
linkage between nodes with respect to node degrees. PTNs of both
large and medium sizes are present in each of the groups. This
indicates the absence of correlations between network size and
degree assortativity $r^{(1)}_{\mathds{L}}$ in $\mathds{L}$-space.
Measuring the same quantity in the $\mathds{P}$ and
$\mathds{C}$-spaces, we observe different behavior. In
$\mathds{P}$-space almost all cities are characterized by very small
(positive or negative) values of $r^{(1)}_{\mathds{P}}$ with the
exception of the PTNs of Istanbul ($r^{(1)}_{\mathds{P}}=-0.12$) and
Los Angeles ($r^{(1)}_{\mathds{P}}=0.12$). On the contrary, in
$\mathds{C}$-space PTNs demonstrate clear assortative mixing with
$r^{(1)}_{\mathds{C}}=0.1\div 0.5$. An exception is the PTN of Paris
with $r^{(1)}_{\mathds{C}}=0.06$.

\begin{table}[ht]
\begin{center}
\tabcolsep1.2mm
\begin{tabular}{lllllll}
\hline\hline
City & $r_{\mathds{L}}^{(1)}$ & $r_{\mathds{L}}^{(2)}$ & $r_{\mathds{P}}^{(1)}$ & $r_{\mathds{P}}^{(2)}$ &
$r_{\mathds{C}}^{(1)}$ & $r_{\mathds{C}}^{(2)}$ \\ \hline
Berlin        &   0.158 & 0.616 &   0.065 & 0.441 & 0.086 & 0.318 \\
Dallas        &   0.150 & 0.712 &   0.154 & 0.728 & 0.290 & 0.550 \\
D\"usseldorf  &   0.083 & 0.650 &   0.041 & 0.494 & 0.244 & 0.180 \\
Hamburg       &   0.297 & 0.697 &   0.087 & 0.551 & 0.246 & 0.605 \\
Hong Kong     &   0.205 & 0.632 &  -0.067 & 0.238 & 0.131 & 0.087 \\
Istanbul      &   0.176 & 0.726 &  -0.124 & 0.378 & 0.282 & 0.505 \\
London        &   0.221 & 0.589 &   0.090 & 0.470 & 0.395 & 0.620 \\
Los Angeles   &   0.240 & 0.728 &   0.124 & 0.500 & 0.465 & 0.753 \\
Moscow        &   0.002 & 0.312 &  -0.041 & 0.296 & 0.208 & 0.011 \\
Paris         &   0.064 & 0.344 &  -0.010 & 0.258 & 0.060 &-0.008 \\
Rome          &   0.237 & 0.719 &   0.044 & 0.525 & 0.384 & 0.619 \\
Sa\~o Paolo     &  -0.018 & 0.437 &  -0.047 & 0.266 & 0.211 & 0.418 \\
Sydney        &   0.154 & 0.642 &   0.077 & 0.608 & 0.458 & 0.424 \\
Taipei        &   0.270 & 0.721 &   0.009 & 0.328 & 0.100 & 0.041 \\
\hline \hline
\end{tabular}
\end{center}
\caption{\label{tab10} Nearest neighbors and next nearest neighbors assortativities
$r^{(1)}$ and $r^{(2)}$ in different spaces  for the whole
PTN. \label{tab3}}
\end{table}

As we have seen above, the PTNs demonstrate assortative
($r^{(1)}>0$) or neutral ($r^{(1)} \sim 0$) mixing with respect to
the node degree (first nearest neighbors number) $k$. Calculating
assortativity with respect to the second next nearest neighbor
number $r^{(2)}$ we explore the correlation of a wider environment
of adjacent nodes. Due to the fact that in this case the two
connected nodes share at least part of this environment (the first
nearest neighbors of a node form part of  the second nearest
neighbors of the adjacent node) one may expect the assortativity
$r^{(2)}$ to be non-negative. Results for $r^{(2)}$ shown in Table
\ref{tab3} appear to confirm this assumption. In all the spaces
considered, we find that all PTNs that belong to the group of
neutral mixing with respect to $k$ also belong to the same group
with respect to the second nearest neighbors. For those PTNs that
display significant nearest neighbors assortativity $r^{(1)}$ we
find that the second nearest neighbor assortativity $r^{(2)}$ is in
general even stronger in line with the above reasoning.

Recall that both closeness and graph centralities ${\cal C}^c$ and
${\cal C}^g$ are measured in terms of path lengths, Eqs.
(\ref{4.9}),  (\ref{4.10}). It is natural to expect that adjacent
nodes will have very similar (or almost identical) centralities
${\cal C}^c$ and ${\cal C}^g$. In turn this will lead to strong
assortative mixing with high assortativities $r^c$ and $r^g$. This
assumption holds only if the average path length in the network is
sufficiently large. The latter is certainly the case for PTNs in
$\mathds{L}$-space but it does not hold in $\mathds{P}$ and even
less in $\mathds{C}$-spaces. Indeed, in $\mathds{L}$-space, where
most PTNs display a mean path length
$\langle\ell_{\mathds{L}}\rangle>10$  (see Table \ref{tab2}) we find
values of $r^c_{\mathds{L}}$ in the range
$r^c_{\mathds{L}}=0.904\div 0.998$ ($r^g_{\mathds{L}}=0.914\div
0.999$). Exceptions are the two PTNs of cities with the smallest
mean paths. These are Moscow ($r^c_{\mathds{L}}=0.865$,
$r^g_{\mathds{L}}=0.870$) with $\langle\ell_{\mathds{L}}\rangle=7.0$
and Paris ($r^c_{\mathds{L}}=0.831$, $r^g_{\mathds{L}}=0.800$) with
$\langle\ell_{\mathds{L}}\rangle=6.4$.

In  $\mathds{P}$ and $\mathds{C}$-spaces where the mean path lengths
are much shorter (of the order of three in $\mathds{P}$ and of the
order of two in $\mathds{C}$-spaces) the one-step difference in path
length between adjacent nodes leads to much reduced assortative
mixing. Numerically this is reflected in much lower (however
positive) values of corresponding assortativities for PTNs where
$\langle\ell \rangle$ is especially small. Indeed, for all PTNs that
display in ${\mathds{P}}$-space a mean path length
$\langle\ell_{\mathds{P}}\rangle < 2.7$ we find
$r^c_{\mathds{P}}<0.5$ ($r^g_{\mathds{P}}<0.4$). At the same time,
PTNs with larger $\langle\ell_{\mathds{P}}\rangle$ may display
larger assortativities even in ${\mathds{P}}$-space. The extreme
example is Los Angeles with $\langle\ell_{\mathds{P}}\rangle = 4.3$
and $r^c_{\mathds{P}} = 0.914$, $r^g_{\mathds{P}}=0.844$. In
$\mathds{C}$-space, where vertices are routes the mean path length
is even smaller and further reduction of closeness and graph
centrality assortativities is observed. For five PTNs we find in
$\mathds{C}$-space $\langle\ell_{\mathds{C}}\rangle<2$ (see Table
\ref{tab2}) and for these $r^c_{\mathds{C}}<0.3$,
$r^g_{\mathds{C}}<0.3$. Again the largest values are attained in the
Los Angeles PTN with
 $\langle\ell_{\mathds{C}}\rangle = 3.4$
and $r^c_{\mathds{C}} = 0.828$, $r^g_{\mathds{C}}=0.648$.

For the other generalized assortativities (stress and betweenness
centrality assortativities $r^s$ and $r^b$ and clustering
coefficient assortativity  $r^{cl}$ we in general find no evidence
for any (positive or negative) correlation in any of the spaces
considered. The only exception are the stress and betweenness
centrality assortativities in ${\mathds{L}}$-space,
$r^s_{\mathds{L}}$ and $r^b_{\mathds{L}}$. There, small but
significantly positive values of $r^s_{\mathds{L}}$ and
$r^b_{\mathds{L}}$ are found. The latter is explained  by the
relatively large mean path length in this space in conjunction with
relatively small node degree values. Let us recall that stress and
betweenness centralities essentially count the number of shortest
paths mediated by a given node. If a selected node is a part of many
such long paths while having low degree, there is  high probability
that any of its neighbors will also be a part of these paths.
Consequently, a positive value of $r^c$ ($r^g$) will arise. The
analogous conclusion can be drawn for nodes with low betweenness (or
stress) centralities. For most PTNs the values of the
assortativities under consideration change in the range
$r^s_{\mathds{L}}=0.26\div 0.64$, $r^b_{\mathds{L}}=0.20 \div 0.61$.
Exceptions are the PTNs which in ${\mathds{L}}$-space have mean path
length $\langle\ell_{\mathds{L}}\rangle < 10$, namely Moscow, Paris
and Sa\~o Paolo. There we find $r^s_{\mathds{L}}=0.02\div 0.10$,
$r^b_{\mathds{L}}=0.02 \div 0.10$.

\begin{figure*}[]
\centerline{
\includegraphics[width=60mm]{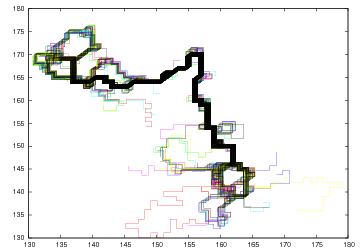}
\includegraphics[width=60mm]{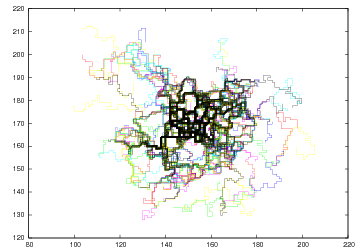}
\includegraphics[width=60mm]{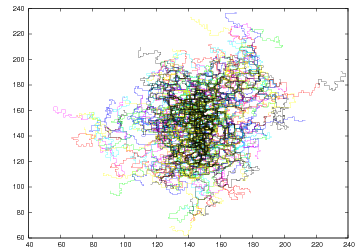}
}
 \centerline{$b=0.1$ \hspace{50mm} $b=0.2$ \hspace{50mm} $b=0.5$}
 \centerline{
\includegraphics[width=60mm]{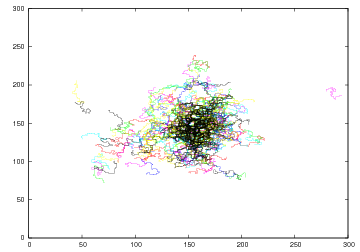}
\includegraphics[width=60mm]{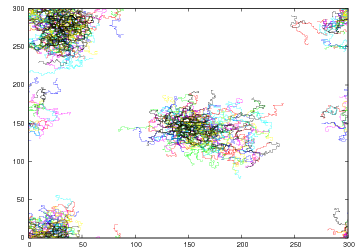}
\includegraphics[width=60mm]{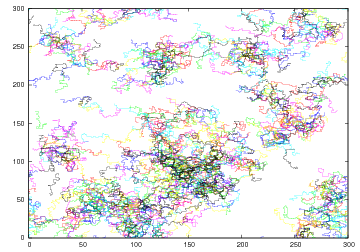}
}
 \centerline{$a=15$ \hspace{50mm} $a=20$ \hspace{50mm} $a=500$}
  \caption{PTN maps of different simulated cities of size $300\times
  300$ with $R = 1024$ routes of $S = 64$ stations each (color online). First row: $a=0$, $b=0.1 \div 0.5$. Second row:
  $b=0.5$, $a=15 \div 500$. With an increase of $b$
  routes cover more and more area.  Increase of $a$ leads to clusterisation of the network.
   \label{fig17}}
\end{figure*}

\section{Modeling PTNs}\label{VI}

\subsection{Motivation and description of the model} \label{VIa}

Having at hand the above described wealth of empirical data and
analysis with respect to typical scenarios found in a variety of
real-world PTNs we feel in the position to propose a model that
albeit being simple may capture the characteristic features of these
networks. Nonetheless it should be capable of discriminating between
some of the various scenarios observed.

If we were only to reproduce the degree distribution of the network,
standard models such as random networks
\cite{Dorogovtsev03,Newman01a} or preferential attachment type
models \cite{prefattach,Amaral00,Liu02,Newman01b,Li03,Ramasco04}
would suffice. The evolution of such networks however is based on
the attachment of nodes. For description of PTNs the concept of
routes as finite sequences of stations is essential
\cite{Holovatch06,Angeloudis06,Ferber07a,Ferber07b} and allows for
the representation with respect to the spaces defined above.
Moreover, taking a route as the essential element of PTN growth
allows to account for the essential bipartite structure of this
network \cite{Seaton04,Zhang06,Guillaume06,Chang07}. Therefore, the
growth dynamics in terms of routes will be a central ingredient of
our model. Another obvious requirement is the embedding of this
model in two-dimensional space. To simplify matters we will restrict
the model to a two-dimensional grid, in particular  to square
lattice. Both the observations of power law degree distributions as
well as the occurrence of the corresponding harness distributions
described above indicate a preference of routes to service common
stations (i.e. an attraction between routes).

Let us describe our model in more detail. As noticed above, a route
will be modeled as a sequence of stations that are adjacent nodes on
a two-dimensional square lattice. Noting that in general loops in
PTN routes are almost absent, a most simple choice to model a PTN
route is a self-avoiding walk (SAW). It may sound less obvious  that
a route apart from being non self-intersecting proceeds randomly.
However, the analysis of geographical data \cite{Ferber07a} has
shown that the fractal dimension of PTN routes closely coincides
with that of a two-dimensional SAW, $d_f=4/3$ \cite{Nienhuis82}. To
incorporate all the above features the model is set up as follows. A
model PTN consists of $R$ routes of $S$ stations each constructed on
a possibly periodic $X\times X$ square lattice. The dynamics of the
route generation adheres to the following rules:
\begin{itemize}
\item 1. Construct the first route as a SAW of $S$ lattice sites.
\item 2. Construct the $R-1$ subsequent routes as SAWs with the following preferential attachment
rules:

{\bf a)} choose a terminal station at $\vec{x}_0$ with probability
\begin{equation} \label{6.1}
p\sim k_{\vec{x}_0} +a/X^2;
\end{equation}

{\bf b)} choose any subsequent station $\vec{x}$ of the route with probability
\begin{equation} \label{6.2}
p\sim k_{\vec{x}} +b.
\end{equation}
\end{itemize}
In (\ref{6.1}), (\ref{6.2}) $k_{\vec{x}}$ is the number of times the
lattice site $\vec{x}$ has been visited before (the number of routes
that pass through $\vec{x}$). Note that to ensure the SAW property
any route that intersects itself is discarded and its construction
is restarted with step 2a).

\begin{table*}[tbh]
\centering
\tabcolsep1.0mm
\begin{tabular}{llllllllllllllllllll}\hline \hline
$R$ & $S$ &  $b$ &   $\langle k_{\mathds{L}} \rangle$    &
$z_{\mathds{L}}$ & $\ell_{\mathds{L}}^{\rm max}$
         & $\langle \ell_{\mathds{L}} \rangle$ & $\langle {\cal C_{\mathds{L}}}^b\rangle$ &
         $\langle k_{\mathds{P}} \rangle$    &  $z_{\mathds{P}}$ &
$\ell_{\mathds{P}}^{\rm max}$
         & $\langle \ell_{\mathds{P}} \rangle$ & $\langle {\cal C_{\mathds{P}}}^b\rangle$ &
         $c_{\mathds{P}}$
&   $\langle k_{\mathds{C}} \rangle$    &  $z_{\mathds{C}}$ &
$\ell_{\mathds{C}}^{\rm max}$
         & $\langle \ell_{\mathds{C}} \rangle$ & $\langle {\cal C_{\mathds{C}}}^b\rangle$ &
         $c_{\mathds{C}}$
         \\ \hline
256 &16 &0.5& 2.92    & 1.66 &    61 & 20.8 &  4.7$\times 10^3$&  44.15 & 3.18 &     7 & 3.0 & 4.7$\times 10^2$   & 7.98 & 86.39 & 1.36 &     6 & 1.9 & 1.2$\times 10^2$ & 2.22 \\
256 &16 &5.0& 2.99    & 1.74 &    80 & 21.7 &  7.5$\times 10^3$&  42.95 & 3.76 &     9 & 3.4 &   8.8$\times 10^2$ & 11.7 & 59.96 & 1.99 &     8 & 2.2 & 1.5$\times 10^2$ & 2.79 \\
256 &32 &0.5& 2.76    & 1.60 &   127 & 38.1 & 3.0$\times 10^4$ &  84.45 & 4.32 &     8 & 3.3 &  1.9$\times 10^3$  & 13.6 & 60.51 & 1.75 &     7 & 2.2 & 1.6$\times 10^2$ & 2.90 \\
256 &32 &5.0& 2.90    & 1.72 &   177 & 43.1 & 5.3$\times 10^4$ &  74.24 & 5.22 &    10 & 4.0 &  3.8$\times 10^3$  & 23.7 & 33.06 & 2.69 &     9 & 2.8 & 2.3$\times 10^2$ & 4.55 \\
512 &16 &0.5& 2.95    & 1.68 &    73 & 22.5 &  6.7$\times 10^3$&  50.07 & 3.39 &     7 & 3.1 & 6.5$\times 10^2$   & 9.14 & 169.7 & 1.44 &     6 & 1.9 & 2.3$\times 10^2$ & 2.25 \\
512 &16 &5.0&  3.12   & 1.78 &    80 & 23.3 & 1.0$\times 10^4$ &  51.56 & 3.79 &    10 & 3.5 &  1.2$\times 10^3$  & 12.3 & 115.3 & 2.24 &     9 & 2.1 &   2.9$\times 10^2$ & 2.88 \\
512 &32 &0.5&  2.83   & 1.63 &   166 & 44.2 & 4.7$\times 10^4$ &  99.53 & 4.56 &    10 &  3.6 &  2.8$\times 10^3$ & 15.7 & 118.4 & 2.03 &     9 & 2.2 & 3.0$\times 10^2$ & 2.92 \\
512 &32 &5.0&  3.12   & 1.79 &   175 & 44.6 & 7.2$\times 10^4$ &  97.05 & 5.37 &     9 & 3.9 &  4.7$\times 10^3$  & 22.2 & 60.36 & 3.08 &     8 & 2.7 & 4.4$\times 10^2$ & 5.04 \\
1024 & 64 & 0.5& 2.86 & 1.66 &   325 & 80.7 & 3.3$\times 10^5$ &  242.2 & 6.32 &     9 & 3.7 & 1.1$\times 10^4$   & 23.4 & 213.3 & 2.42 &     8 & 2.2 & 6.1$\times 10^2$ & 3.10 \\
1024 & 64 & 1.0& 2.97 & 1.72 &   355 & 88.5 & 4.8$\times 10^5$ &  222.2 & 6.74 &    12 & 4.2 & 1.7$\times 10^4$   & 32.4 & 143.9 & 2.97 &    11 & 2.5 & 7.9$\times 10^2$ & 4.39 \\
\hline \hline
\end{tabular}
\caption{Characteristics of the simulated PTN with $X=300$, $a=0$
for different parameters $R$, $S$,  and $b$. The rest of notations as in Table
\ref{tab2}. \label{tab5}}
\end{table*}

\subsection{Global topology of model PTN} \label{VIb}

Let us first investigate the global topology of this model as
function of its parameters. We first fix both the number of routes
$R$ and the number of stations $S$ per route as well as the size of
the lattice $X$. This leaves us with essentially two parameters $a$
and $b$, Eqs. (\ref{6.1}), (\ref{6.2}). Dependencies on $R$, $S$,
and $X$ will be studied below.

For the real-world PTNs as  studied in the previous sections, almost
all stations belong to a single component, GCC, with the possible
exception of a very small number of routes. Within the network
however we often observe what above we  called the harness effect of
several routes proceeding in parallel for a sequence of stations.
Let us first investigate from a global point of view which
parameters $a$ and $b$ reproduce realistic maps of PTNs. In Fig.
\ref{fig17} we show simulated PTNs on lattices $300\times 300$ for
$R=1024$, $S=64$ and different values of the parameters $a$ and $b$.
Each route is represented by a continuous line tracing the path
along its sequence of stations. For representation purposes,
parallel routes are shown slightly shifted. Thus, the line thickness
and intensity of colors indicate the density of the routes.

Parameter $b$ governs the evolution of each single subsequent route.
If $b=0$ each subsequent route is restricted to follow the previous
one. The change of simulated PTNs with $b$ for fixed $a=0$ is shown
in the first row of Fig. \ref{fig17}. For small values of $b=0\div
0.1$ the PTNs obtained result in almost all routes following the
same path with only a few deviations. Increasing $b$ from $b=0.1$ to
$b=0.2$ the area covered by the routes increases while the majority
of the routes are concentrated on a small number of paths. Further
increasing $b$ to $b=0.5$ and beyond we find a wider distributed
coverage with the central part of the network remaining the most
densely covered area. This is due to the non-equilibrium growth
process described by Eqs. (\ref{6.1}), (\ref{6.2}).

The parameter $a$ quantifies the possibility to start a new route
outside the existing network. For vanishing $a=0$ the resulting
network always consists of a single connected component, while
for finite values of $a$ a few or many disconnected components may occur. The
results found for $a=0$ and varying $b$ parameters are completely
independent of the lattice size $X$ provided $X$ is sufficiently
large. When introducing a finite $a$ parameter, however, new routes may be
started anywhere on the lattice which results in a strong lattice
size dependency. To partly compensate for this, the impact of $a$
has been normalized by $X^2$ in (\ref{6.1}). The dependence of the
simulated PTN maps on $a$ for fixed $b=0.5$ is shown in the second
row of Fig. \ref{fig17}. For $a< 15$ one observes the formation of a
single large cluster with only a few individual routes occurring
outside this cluster. Slightly increasing $a$ beyond $a=15$ one
finds a sharp transition to a situation with several (two or more)
clusters. For much larger values of $a$ the number of clusters
further increases and the situation becomes more and more
homogeneous: the routes tend to cover all available lattice space
area.

\subsection{Statistical characteristics of model PTN} \label{VIc}


\begin{figure*}
\centerline{
\includegraphics[width=55mm]{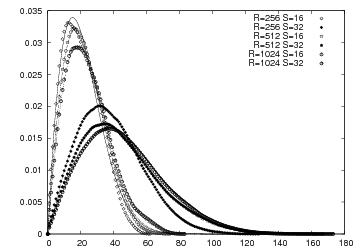}
\includegraphics[width=55mm]{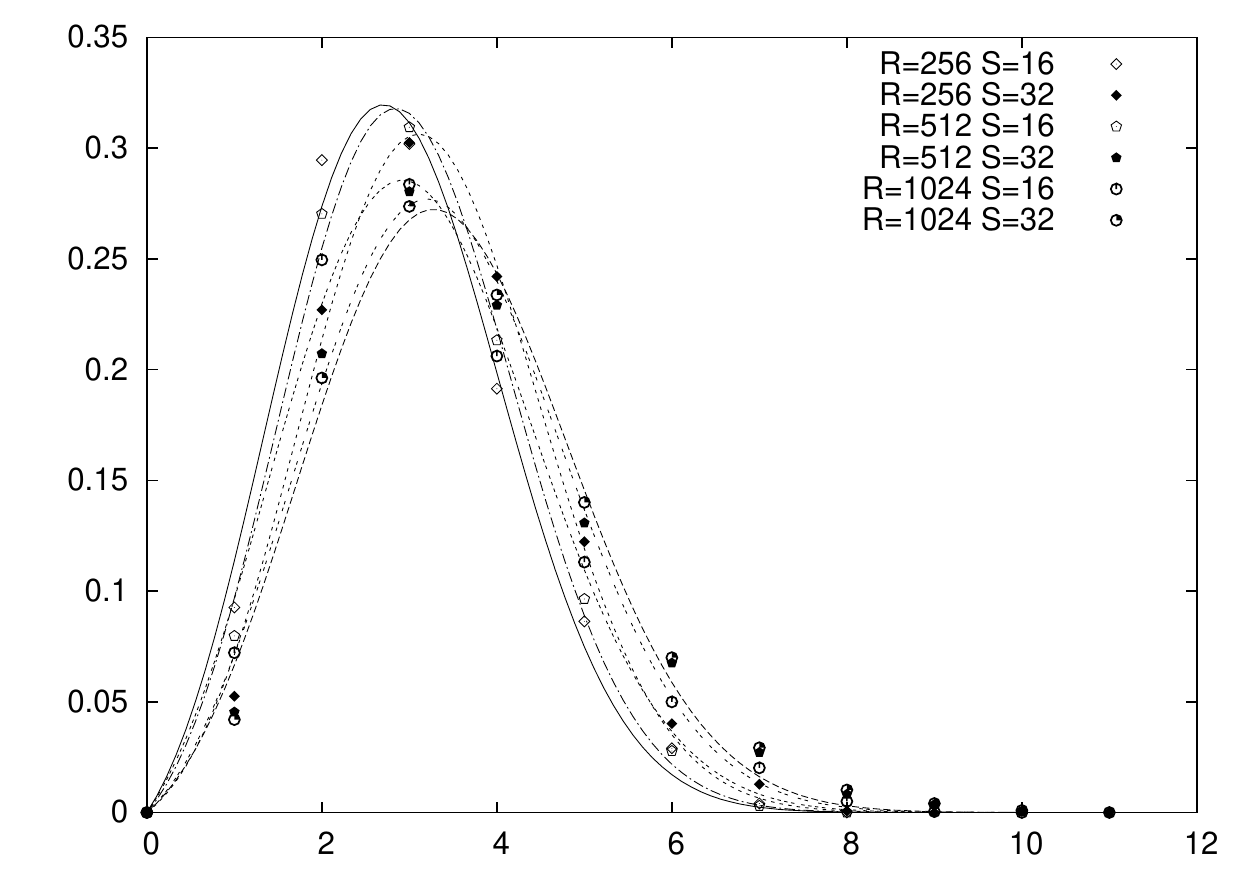}
\includegraphics[width=55mm]{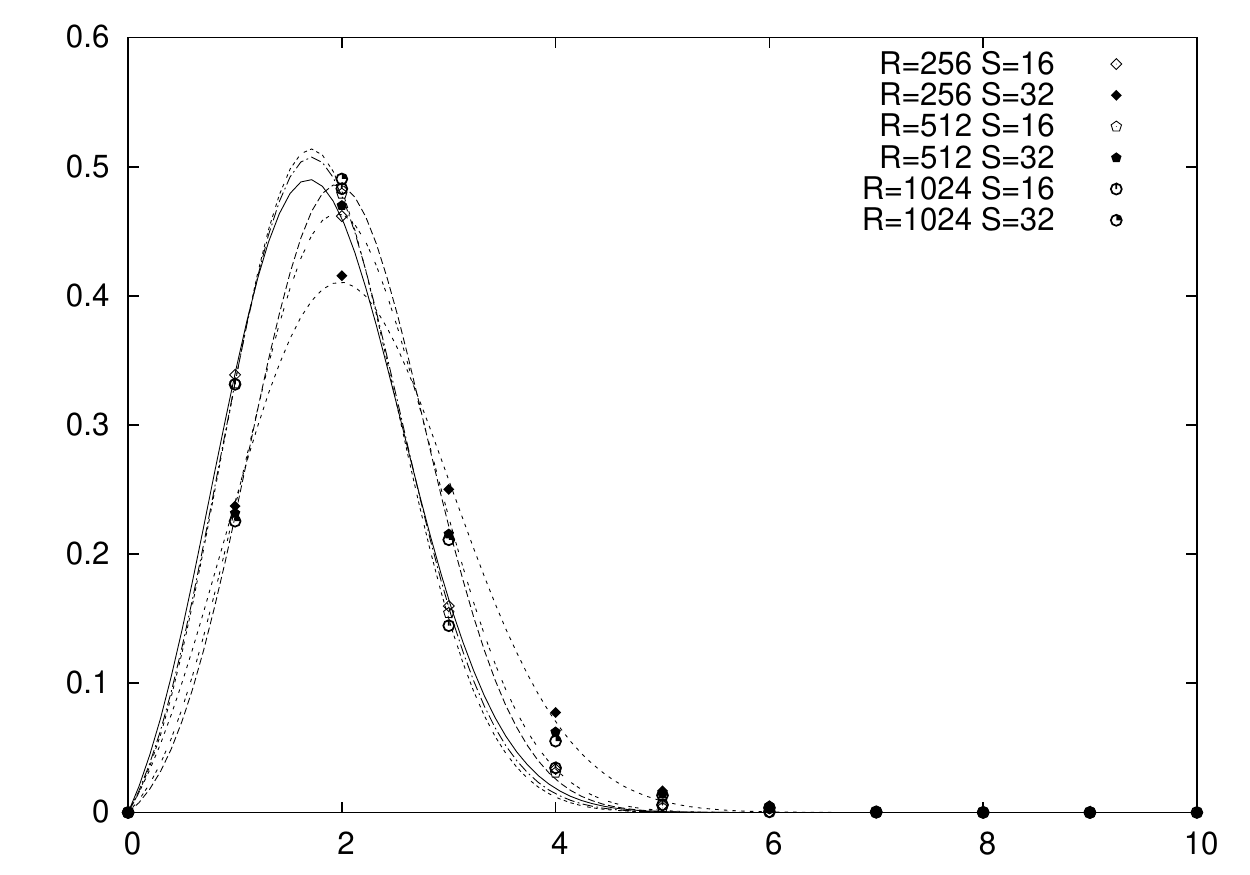}
 }
 \centerline{
{\bf a} \hspace{17em} {\bf b} \hspace{17em} {\bf c}
 }
\caption{Mean shortest path length distribution $P(\ell)$ for
several simulated PTNs. {\bf a}:
$\mathds{L}$-space; {\bf b}: $\mathds{P}$-space; {\bf c}: $\mathds{C}$-space.
Symbols correspond to simulation results, curves to fits of unimodal distributions.}
 \label{fig18}
\end{figure*}

\begin{figure*}
\centerline{
\includegraphics[width=85mm]{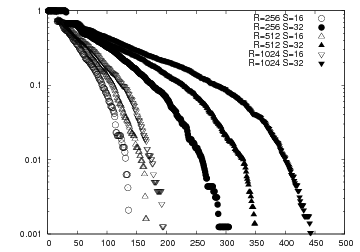}
\includegraphics[width=85mm]{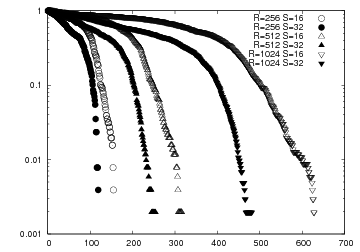}}
\centerline{{\bf a} \hspace{19em} {\bf b}}
 \caption{Cumulative node degree
 distributions $P^{\rm cum}(k)$ (\ref{3.2a})
 for several simulated PTNs in ({\bf a}) $\mathds{P}$  and  ({\bf b}) $\mathds{C}$-spaces.}
 \label{fig19}
\end{figure*}


From the above qualitative investigation we conclude that realistic
PTN maps are obtained for small or vanishing $a$ and $b\geq 0.5$. To
quantitatively investigate the behavior of the simulated networks
on their parameters including $R$ and $S$ let us now compare their
statistical characteristics with those we have empirically obtained
for the real-world networks. In Table \ref{tab5} we have chosen to
list the same characteristics of the simulated PTN as they are
displayed for the real-world networks in Table \ref{tab2}. To
provide for additional checks of the correlations between simulated
and real-world networks, we present the characteristics in all
$\mathds{L}$, $\mathds{P}$, and $\mathds{C}$-spaces. Let us note
that our choice of the underlying grid to be a square lattice limits
the number of nearest neighbors of a given station in
$\mathds{L}$-space to $k_{\mathds{L}}\leq 4$. Moreover, as far as no
direct links between these neighbors occur, the clustering
coefficient in $\mathds{L}$-space vanishes, $c_{\mathds{L}}=0$.
Nonetheless, as we discuss below, both characteristics display
nontrivial behavior similar to real-world networks when
displayed in $\mathds{P}$ and $\mathds{C}$-spaces.

For reasons explained above we choose a vanishing parameter
$a=0$ and $b=0.5$ and for comparison $b=5.0$. The data shown in the Table was
obtained for simulated PTNs of different numbers of routes, $R=256,
\, 512, \, 1024$ and route lengths $L=16, \, 32, \, 64$. In the
range of parameters covered in the Table we observe only weak
changes of the various characteristics. Natural trends are that with the
increase of the number of routes $R$  the maximal and mean shortest path length increases
in all spaces. Most pronounced this is observed in
$\mathds{L}$-space, while it is weakest  in $\mathds{C}$-space. A
similar increase is observed in $\mathds{L}$-space when increasing
the number of stations $S$ per route. Choosing the values of $R$ in
the range $R=256\div 1024$ and $S=16$, $S=32$ the average and
maximal values of the characteristics studied here are found within
the ranges seen for real-world PTNs, see Table \ref{tab2}. More
detailed information is contained in the distributions of these
characteristics and their correlations.

We restrict the further discussion to simulated PTNs described by
$R=256, 512, 1024$, $S=16,32$, and $a=0$, $b=0.5$, which appear to
reproduce many of the characteristics of real-world PTNs. In figure
\ref{fig18} we display the mean shortest path length distribution
for these selected PTNs in $\mathds{L}$, $\mathds{P}$, and
$\mathds{C}$-spaces. In $\mathds{L}$-space we observe two groups of
distributions which correspond to the two route lengths $S=16$ and
$S=32$. The most probable values for the path length
$\hat{\ell}_{\mathds{L}}$ being of the order of the corresponding
$S$. In $\mathds{P}$ and $\mathds{C}$-spaces the distributions are
very similar with most probable path lengths
$\hat{\ell}_{\mathds{P}}\sim 3$, $\hat{\ell}_{\mathds{C}}\sim 2$. In
all cases the distributions are well fitted by the asymmetric
unimodal distribution (\ref{4.3}) and resemble those of the
real-world networks shown in Fig. \ref{fig7}. Varying  $b=0.2\div 5$
does not significantly change this picture.

\begin{figure*}[]
\centerline{
\includegraphics[width=65mm,angle=0]{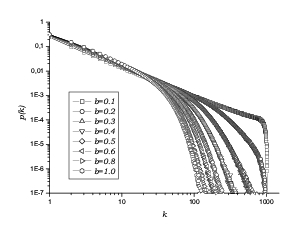}
\includegraphics[width=65mm,angle=0]{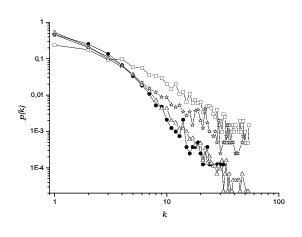}
}
 \centerline{
{\bf a} \hspace{17em}  {\bf b}
 }
\caption{{\bf a}: Averaged (over $3000 \div 30000$ simulated cities)
station node degree
   distributions in $\mathds{B}$-space. $R=1024$, $S=64$, $a=0$. Parameter
$b$ changes in the region
   $b=0.1 \div 1.0$ as shown in the legend. {\bf b}: Corresponding node
degree distributions  for
   Hamburg (circles), Hong Kong (squares), Los Angeles (triangles), and
Istanbul (stars).  \label{fig20}}
\end{figure*}

\begin{figure*}
\centerline{
\includegraphics[width=55mm]{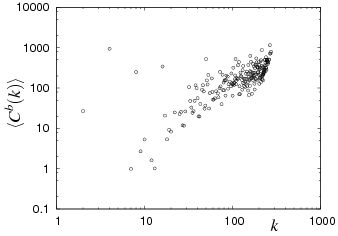}
\includegraphics[width=55mm]{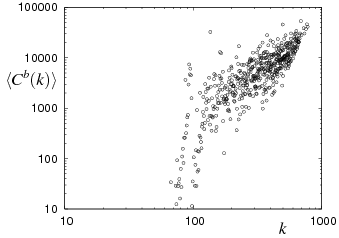}
\includegraphics[width=55mm]{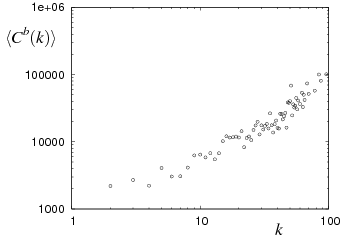}}
\centerline{ {\bf a} \hspace{17em} {\bf b} \hspace{17em} {\bf c}
 }
\caption{Mean betweenness centrality $\langle {\cal C}^b(k)\rangle$
for the simulated city of
$300\times 300$ sites with $R=500$, $S=50$, $a=0$, and  $b=0.5$ in $\mathds{C}$ ({\bf a}), $\mathds{P}$
({\bf b}),
and $\mathds{B}$ ({\bf c}) spaces. }
 \label{fig21}
\end{figure*}

Let us now examine the node degree distributions of the simulated
PTNs selected above. As explained above, the $\mathds{L}$-space
degrees are restricted by the geometry of the underlying square
lattice. Thus of the representations discussed here one may observe
non-trivial distributions only in $\mathds{P}$, $\mathds{C}$, and
$\mathds{B}$-spaces. Fig. \ref{fig19}{\bf a} shows the cumulative
node degree distribution in $\mathds{P}$-space in semi-logarithmic
scale. Recall that for the majority of real-world PTNs studied in
section \ref{III} as well as in other works
\cite{Sienkiewicz05,Xu07}, the $\mathds{P}$-space node degree
distribution was found to decay exponentially. All distributions
shown in Fig. \ref{fig19}{\bf a} display two regions each governed by an
exponential decay with a separate scale. Note that increasing
both $S$ and $R$ leads to an increase of the ranges over which these
regions extend. Comparing these results with those of Fig.
\ref{fig4}{\bf b} for real-world PTNs we find that all ranges
observed there are also reproduced here. Within the parameter ranges
chosen here the current model does not seem to attain a power law
node degree distribution in $\mathds{P}$-space.

Comparing the$\mathds{C}$-space node degree distributions for real-world and simulated
PTNs (Figs. \ref{fig4}{\bf c} and \ref{fig19}{\bf b},
correspondingly) one finds a definite tendency to an exponential
behavior with two different scales in both cases. Note however that
for the simulated PTNs the scales increase with the number of routes
$R$ while they decrease with the number of stations per route $S$.

The simulated results discussed so far concerned data obtained for
individual instances of modeled PTNs. One of the reasons for this
was to reduce the computational effort required for the calculation
of path lengths, betweennesses, and related global characteristics.
Furthermore, in particular for the simulations involving high number
of routes some self averaging may be expected to occur. The latter
assumption was tested and verified by (i) simulating a reasonable
set of PTNs with the same choice of parameters and (ii) by
performing large-scale simulations calculating local
characteristics. A result of  the latter procedure involving
averages over up to $3\cdot 10^4$ instances of simulated networks is
shown in Fig. \ref{fig20}{\bf a}. There we show the node degree
distribution of the station nodes in $\mathds{B}$-space, i.e. the
bipartite network of routes and stations with the inherent
neighborhood relation (see Fig. \ref{fig3}). As can be seen in the
double logarithmic plot shown in Fig. \ref{fig20}{\bf a} a power-law
like behavior of this distribution that extends over a wider range
is found for small values of the parameter $b$. This corresponds to
a situation where one finds many routes to proceed in parallel
(compare with the maps shown in Fig. \ref{fig17}). For the more
realistic choices of the $b$ parameter the overall behavior of this
distribution is described by an exponential decay. The scale of this
decay strongly depends on $b$. Fig. \ref{fig20}{\bf b} shows that
similar distributions for the real cities have oscillating
character, which is caused by the fact that non-cumulative
distributions are plotted. Similarly, individual distributions for
simulated PTNs are in general non-monotonous, however the large
number average of the distribution appears to be monotonously
decreasing. Nevertheless, comparing plots in Figs. \ref{fig20}{\bf
a} and \ref{fig20}{\bf b} one sees that in general the model is
capable to reproduce the global decay properties of the station node
degree distributions in $\mathds{B}$-space.

In Fig. \ref{fig21} we show the betweenness-degree correlation for
the simulated PTN with $X=300$, $R=500$, $S=50$, $a=0$, and $b=0.5$.
There, we present the mean betweenness centrality $\langle {\cal
C}^b(k)\rangle$ in $\mathds{C}$, $\mathds{P}$, and
$\mathds{B}$-spaces. Corresponding plots for a real world network
are shown in Fig. \ref{fig14}. Plots displayed for the simulated
networks in Figs. \ref{fig21}{\bf a} - \ref{fig21}{\bf c}
qualitatively reproduce the behavior of $\langle {\cal
C}^b(k)\rangle$ observed for the real world networks in
$\mathds{C}$, $\mathds{P}$, and $\mathds{B}$-spaces.
$\mathds{L}$-space behavior can not be reproduced due to the
restrictions caused by the geometry of the underlying square
lattice.

\begin{figure}
\centerline{
\includegraphics[width=45mm]{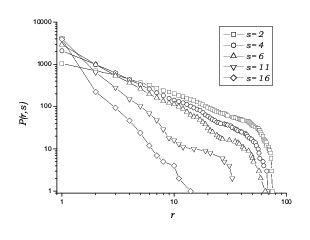}
\includegraphics[width=45mm]{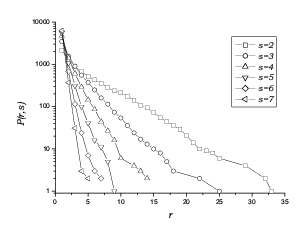}}
\centerline{{\bf a} \hspace{17em} \bf{b} }
 \caption{Cumulative harness distributions $P(r,s)$ for the simulated PTN with
$R=256$, $S=32$. Fig. {\bf a}:  $a=0$, $b=0.2$, fig. {\bf b}: $a=0$, $b=1.0$.
Compare with plots in Figs. \ref{fig15}, \ref{fig16} for the
real-world networks.}
 \label{fig22}
\end{figure}

In Figs. \ref{fig22}{\bf a} and \ref{fig22}{\bf b} we plot the
cumulative harness distributions $P(r,s)$ for two simulated networks
with $R=256$, $S=32$, $a=0$ and different values of parameter $b$:
$b=0.2$ (Fig. \ref{fig22}{\bf a}) and $b=1.0$ (Fig. \ref{fig22}{\bf
b}). Similar plots for real world networks are given in Figs.
\ref{fig15} and \ref{fig16}. The plots of Fig. \ref{fig22} nicely
reproduce two regimes empirically observed for the real-world PTN.
In the first, the harness distribution is governed by a power law
decay (\ref{4.12a}), Fig. \ref{fig22}{\bf a}, whereas in the other
one there is a tendency to an exponential decay (\ref{4.13}), Fig.
\ref{fig22}{\bf b}. A prominent feature demonstrated by Fig.
\ref{fig22} is that one can tune the decay regime by changing the
parameter $b$. For small values of $b$ the probability of a
route to proceed in parallel with other routes is high c.f. Eq.
(\ref{6.2}). Therefore, the number of ``hubs'' in the $P(r,s)$
distribution of lines of several routes that go in parallel is large
for small $b$. This is reflected by a power-law decay of the
distribution. Alternatively, an increase of $b$ leads to a decrease
of such hubs as shown by the exponential decay of their distribution.

Summarizing the comparison of the statistical characteristics of
real world networks with those of simulated ones one can definitely
state that the model proposed above captures many essential features
of real world PTNs. This is especially evident if one includes into the
the comparison different network representations (different spaces) as
performed  above.

\section{Conclusions}\label{VII}

This paper was driven by two main objectives towards the analysis of
urban public transport networks. First, we wanted to present a
comprehensive survey of statistical properties of PTNs based on the
data for cities of so far unexplored network size (see Table
\ref{tab1}). Based on this survey, the second objective was to present
a model that albeit being simple enough is capable to reproduce a
majority of these properties.

Especially helpful in our analysis was the use of different network
representations (different spaces, introduced in section \ref{II}).
Whereas former PTN studies used some of these representations, here
within a comprehensive approach we calculate PTN characteristics as
they show up in $\mathds{L}$, $\mathds{P}$, $\mathds{C}$, and
$\mathds{B}$-spaces. It is the comparative analysis of empirical
data in different spaces that enabled, in particular, an adequate
PTN modeling presented in section \ref{VI}.

The networks under consideration appear to be strongly correlated
small-world structures with high values of clustering coefficients
(especially in $\mathds{L}$ and less in $\mathds{C}$-spaces) and
comparatively low mean shortest path values, as listed in Table
\ref{tab2}. Standard network characteristics listed there correspond
to the features a passenger is interested in when using public
traffic in a given city. To give several examples, any two stops in
Paris are on the average separated by $\langle \ell_{\mathds{L}}
\rangle-1=5.4$ stations (with a maximal value of
$\ell_{\mathds{L}}^{\rm max}-1=27$) and to travel between them one
on average should do $\langle \ell_{\mathds{P}} \rangle-1=1.7$
changes. Evidence of correlations present in PTNs are the power-law
node degree distributions observed for many networks in $\mathds{L}$
and for some in $\mathds{P}$-space (see Table \ref{tab3}).
Currently, we find no explanation why some of the networks of our
survey are governed by power-law node degree distributions whereas
others follow an exponential decay. In the analysis of
urban street networks a classification has been found
\cite{Cardillo06,Volchenkov} that allows to discriminate between
properties of different classes of city organization. Let us note however that
as a rule the latter analysis is performed for restricted regions of street
networks i.e. either the historical or the suburban part. In the case of a PTN, however, one usually deals with a structure
that  spreads over all the city, covering both the inner and outer regions.

Besides looking on traditional network characteristics (as described
in sections \ref{III} - \ref{V}) we addressed here a specific
feature which is unique for PTNs and networks with similar
construction principles. Namely, we analyzed statistical
distributions of public transport routes that go in parallel for a
sequence of stations. As we have shown such distributions (we call
them harness distributions) are well defined for the networks under
consideration and may be also be used for a quantitative description of
similar networks embedded in 2D or 3D space as cables,
pipes,  neurons, or (blood-) vessels, etc.

The common statistical features of the networks considered emerge
due to their common functional purposes and construction principles also reflected in
the underlying bipartite structure \cite{Guillaume06}.
It is this structure that explains parts of the
correlations present in PTNs \cite{Seaton04}. The network growth
model we present in Section \ref{VI} captures this structure
describing network evolution in terms of adding  public transport
routes, each of them being a complete graph in $\mathds{P}$-space.
Our choice to use a self avoiding walk (SAW) as a route model in
lattice simulations was motivated by geographical observations and other
reasons, as argued in section \ref{VI}. In support of the scaling
argument given there, one may note that the fractal dimension
of a SAW on a lattice does not change if a weak uncorrelated
disorder is present, i.e. when some lattice sites can not be visited
\cite{saw_disorder}. In turn, this tells that the model is robust
with respect to  weak disturbances of the underlying lattice
structure. Further analysis of simulated PTNs performed in section
\ref{VI} established strong similarities in the statistical
characteristics of simulated and real-world networks.

Obviously, the two objectives in the PTN study we have so far achieved in
this paper - the empirical analysis and the modeling - naturally
call for an analytic approach. In particular, such approach may be
used in parallel with numerical simulations to derive statistical
properties of the model proposed in section \ref{VI}. This will be a
task for forthcoming studies. Another natural continuation of this
work will be to analyze different possibly dynamic phenomena that
may occur on and with PTNs. A particular task will be to study
robustness of PTNs to targeted attacks and random failures
\cite{Ferber07c}.

Yu.H. acknowledges support of the Austrian FWF project 19583-PHY.


\begin{thebibliography}{50}
\bibitem{database2} Some numbers for the real-world PTNs  slightly differ from those from
our letter \cite{Ferber07a}. The reason is an improvement of the database.
\end{thebibliography}
\end{document}